\documentclass{article}

\usepackage{jcappub} 
\usepackage[utf8]{inputenc}
\usepackage{appendix}
\usepackage{subfigure}
\usepackage{physics}
\usepackage[sort&compress]{natbib}
\title{Density and magnetic intensity dependence of radio pulses induced by energetic air showers}
\author{Juan Ammerman-Yebra,}
\author{Jaime Alvarez-Mu\~niz,}
\author{Enrique Zas}

\affiliation{Instituto Galego de F\'\i sica de Altas Enerx\'\i as (IGFAE),
Universidade de Santiago de Compostela, 15782 Santiago
de Compostela, Spain}

\emailAdd{juan.ammerman.yebra@usc.es}
\emailAdd{jaime.alvarez@usc.es}
\emailAdd{enrique.zas@usc.es}

\date{\today}

\begin{document}
\abstract{We have studied the effect of changing the density and magnetic field strength in the coherent pulses that are emitted as energetic showers develop in the atmosphere. For this purpose we have developed an extension of ZHS, a program to calculate coherent radio pulses from electromagnetic showers in homogeneous media, to account for the Lorentz force due to a magnetic field. This makes it possible to perform quite realistic simulations of radio pulses from air showers in a medium similar to the atmosphere but without variations of density with altitude. The effects of independently changing the density, the refractive index and the magnetic field strength are studied in the frequency domain for observers in the Cherenkov direction at far distances from the shower. This approach is particularly enlightening providing an explanation of the spectral behavior of the induced electric field in terms of shower development parameters. More importantly, it clearly displays the complex scaling properties of the pulses as density and magnetic field intensity are varied. The usually assumed linear behavior of electric field amplitude with magnetic field intensity is shown to hold up to a given magnetic field strength at which the extra time delays due to the deflection in the magnetic field break it. Scaling properties of the pulses are obtained as the density of air decreases relative to sea level. A remarkably accurate scaling law is obtained that relates the spectra of pulses obtained when reducing the density and increasing the magnetic field.}

\maketitle

\section{Introduction}

The detection of high-energy cosmic ray showers using the radio technique has been explored for a long time, with radio pulses from cosmic ray showers having been detected in the atmosphere since the late 1950s'~\cite{JELLEY1965}. Experimental activity was reduced in the 1970s', but the technique was revived in the 1990s' as an alternative for detecting High-Energy (HE) neutrinos interacting in Antarctic ice \cite{Gusev:1983wuo}.  
The radio technique experienced a significant surge at the turn of the XXI century with new experiments using fast electronics and powerful recording systems, some combining radio detection with conventional air shower detectors~\cite{Miocinovic2005,Rottgering2003,Fuchs2012,Bezyazeekov2015} to study pulses induced by cosmic rays. Efforts to calculate pulses in showers crystallized into reliable predictions when the first program capable of calculating the coherent electromagnetic emission was developed, the Zas-Halzen-Stanev code, often referred to as ZHS~\cite{zas1992electromagnetic}. Later, in 2012, the approach used for ice~\cite{zas1992electromagnetic,Alvarez-Muniz:2010wjm,Alvarez-Muniz:2022uey} and similar ones~\cite{James:2010vm} were applied to well-established air shower simulation programs that constitute current standards for detailed calculations~\cite{AlvarezMuniz2012,Huege:2013vt}. These programs pointed to the transverse currents induced by the magnetic field of the Earth as the main mechanism responsible for the coherent radio emission in air showers, confirming earlier  work~\cite{kahn1966radiation,Clay:1969}. 
Some authors also invoke synchrotron emission, particularly at the highest frequencies~\cite{James:2022mea,Huege:2003up, chicheARENA2022}. The enormous progress that followed in the last decade has revealed an ever increasing potential of the radio technique in air, with very interesting applications for the study of both Ultra-High-Energy (UHE) Cosmic Rays and neutrinos, leading to a multitude of experimental initiatives~\cite{Schroder:2016hrv,Connolly:2016pqr,Huege:2017khw}. 

Progress in the description of the emission patterns of the coherent radio pulses in the atmosphere, both for the design and the analyses of different experiments, is based on dedicated and extensive simulations to cover the relevant parameter space of primary energy, zenith and azimuth angles, and observer positions (typically arranged in a large grid). This is very resource consuming. Most valuable information about pulses in air showers has been obtained with simulations for given experimental situations that, inevitably, involve a diversity of conditions in which the pulses are obtained for different experimental setups. The conditions may actually change within a single experiment. For instance, for an array of antennas measuring cosmic rays, as the zenith angle increases, the showers develop in a lower density atmosphere and become significantly larger, while at the same time the Cherenkov angle in the region of shower maximum diminishes to angles well below half of a degree. Simultaneously, the angles between the magnetic field and the shower direction will change as the zenith and azimuth angles change. Other experimental choices such as the precise location of the detector in the Earth, the frequency band, or the altitude of the detector will further modify the pulses. 

There are no simple models to describe the pulses in arbitrary circumstances, complicating the design and optimization of experiments and the reconstruction and interpretation of data taken. Efforts have been made to finding shortcut approaches based on interpolation, models or parameterizations that reduce the number of observation points to calculate~\cite{Buniy:2000kk, Scholten:2007ky, Werner:2007kh,Alvarez-Muniz:2011wcg, Werner:2012cr, Alvarez-Muniz:2020ary, Zilles:2018kwq, Tueros:2020buc}. These would benefit from good understanding of the general properties of the radio pulses. In air, the separation of the excess charge and geomagnetic effects, each responsible for a polarization pattern with very different symmetry, has been particularly useful in this respect~\cite{Alvarez-Muniz:2014wna,Glaser:2018byo,Zilles:2018kwq,Tueros:2020buc, Schluter:2022mhq}. 
It is generally assumed that the amplitude of the pulses scales with the projection of the magnetic field intensity in the plane perpendicular to shower direction as the Lorentz force is proportional to it~\cite{PierreAuger:2016vya, falcke2005detection, aab2014probing, ardouin2009geomagnetic, T-510:2015pyu}. 
The amplitude of the pulses in air has been suggested to behave in a well-defined manner with the air density~\cite{Scholten:2007ky,deVries:2011pa,T-510:2015pyu}, but simulations for up-going showers from Earth-skimming tau neutrinos hint to deviations from simple scaling laws~\cite{Romero-Wolf:2021D+}, an effect also seen in simulations of very inclined showers~\cite{Chiche:2021iin, chicheARENA2022,Schluter:2022mhq}. 
The combined effects of air density and geomagnetic field intensity are however not easy to disentangle since they can take place simultaneously in the same atmospheric shower and depend on shower direction and location of the observers in a given experiment. For this reason it is difficult to obtain generalities from detailed simulations of very specific experimental conditions. 
It is not surprising that there are no comprehensive descriptions in the literature that fully address radio emission in air and all its dependencies. It seems a daunting task to take all these effects into consideration to describe how the pulses will change under different circumstances. 

In this work we attempt to shed light on the influence of air density and magnetic field on the radio emission in atmospheric showers using simulations performed with the ZHS code ~\cite{zas1992electromagnetic}. At a given observer position, ZHS coherently adds the contributions to the electric field produced by charged particles in an electromagnetic shower, dividing their trajectories in tracks which are assumed to be rectilinear and traveled at constant speed. The ZHS code, originally conceived to describe pulses in homogeneous ice, was extended to deal with other homogeneous media including air~\cite{AlvarezMuniz2009}, but it is limited to showers induced by photons or electrons, developing in constant density and ignoring magnetic deflections. As the interaction length in ice is typically very short, the deviations in the magnetic field of the Earth contribute very little to the pulse. Despite these limitations, the ZHS code has had an important role testing other simulation programs of radio pulses~\cite{AlvarezMuniz2012a}, and more importantly because it is presently the only program to simulate coherent transition radiation from ice to air~\cite{Motloch:2015wca}.

The role of the ZHS code in studying radio pulses in the atmosphere has thus been limited to describing the behavior of pulses due to charge excess alone (not accounting for magnetic deflections that are responsible for the bulk of radio emission in air in many experimental situations). In this work, we present an upgraded version of the ZHS simulation code for electromagnetic showers to account for magnetic deflections. While this is no substitute of realistic simulations in the atmosphere, it provides a useful tool to explore the nature of coherent pulses in air showers in simplified conditions. These simulations can help disentangle different effects that often take place simultaneously when pulses are simulated with standard codes that have a realistic description of the Earth's atmosphere. 
We use the modified ZHS program here to study the effect of the air density, the magnetic field strength and the refractive index in the frequency spectra of the radio pulses. We obtain frequency-dependent scaling relations of the pulses as these simulation conditions are independently changed for observers located at large distances, primarily in the Cherenkov direction where the intensity of the radio emission is strongest.
We also find frequency-dependent deviations from simple linear behavior of the spectra as the density is lowered, and a similar effect as the magnetic field intensity is increased. We demonstrate that this is mainly due to the additional time delays induced by magnetic deflections. These findings give important insights into the radio emission from air showers. In addition, the ZHS code with magnetic deflections can also be of interest to make simulations of experimental arrangements that measure the effect of transverse currents in homogeneous media with magnetic fields~\cite{T-510:2015pyu,Bechtol:2021tyd}, and the results obtained here can be relevant for the interpretation of the data taken.

\section{Simulation of radio emission with the ZHS code}

The ZHS program is a Monte Carlo simulation of electromagnetic showers in homogeneous media specifically designed to calculate the properties of the emitted radio pulses~\cite{zas1992electromagnetic}. The code accounts for electromagnetic processes such as bremsstrahlung, pair production, and interactions with matter electrons, namely, M\o ller, Bhabha, Compton scattering and electron-positron annihilation. These interactions produce the Askaryan effect, that is, the development of a negative excess charge as matter electrons are entrained in the shower \cite{Askaryan:1962}, which is mostly responsible for the radio emission in a dense medium. Multiple elastic scattering (using  Moli\`ere's theory) and ionization losses are implemented as continuous effects. The ZHS code was originally designed to follow all electrons and positrons in ice down to a $\sim$100 keV kinetic energy threshold and to carefully account for the timing of each particle crucial for interference effects. 
Particle time delays are defined relative to a plane front perpendicular to the shower axis and traveling along this axis at the speed of light. They are due to the propagation geometry and to the sub-luminal particle velocities and they are accounted for assuming uniform energy loss in each step of propagation. An approximate account is also made of the time delay associated to the multiple elastic scattering processes along each step. The effect of deviations of the particles in the magnetic field of the Earth are not relevant for the calculation of radio pulses in a dense medium and the original ZHS program did not account for them. Further details can be found in~\cite{zas1992electromagnetic}. 

The purpose of the ZHS code was to simulate coherent pulses in ice and this was achieved reducing the shower to a superposition of electron and positron tracks assumed to be rectilinear and uniform. Tracks are naturally split in the Monte Carlo by the particle interactions and they are subdivided so that their length is always below 0.1 radiation lengths. For low energies, sub-tracks are further reduced so that their length is always less than a small fraction of the particle range. Ionization losses and multiple elastic scattering are evaluated for each track or step in the propagation. Convergence of results as the step is reduced has been carefully checked \cite{Alvarez:1995,Alvarez-Muniz:2000aah}. The program has been extended to deal with other homogeneous  media~\cite{AlvarezMuniz2009}. 

The original calculation of radio emission was made in the frequency domain, with the Fourier transform of the time-domain electric field $\boldsymbol{E}(t, \boldsymbol{x})$ given by\footnote{Note that we use a non-standard convention for the Fourier transform following \cite{zas1992electromagnetic}.}:
\begin{equation}
	\boldsymbol{E}(\omega, \boldsymbol{x}) = 2 \int_{-\infty}^\infty \boldsymbol{E}(t,
	\boldsymbol{x}) e^{i \omega t} dt 
\end{equation}
The Fourier transform of the electric field radiated by a charged particle moving with uniform speed between two fixed points can be obtained in the Fraunhofer limit from Maxwell's equations. If the velocity is $\boldsymbol{v}$ and the points correspond to times $t_1$ and $t_2 = t_1 + \delta t$, the spectral amplitude of the electric field is given by the expression~\cite{zas1992electromagnetic}:
\begin{equation}
	\label{eq:ZHS}
	\boldsymbol{E}(\omega, \boldsymbol{x}) = \frac{e \mu_r}{2 \pi \epsilon_0 c^2} i \omega \frac{e^{i k R}}{R}
	e^{i(\omega t_1 - \boldsymbol{k} \cdot \boldsymbol{x_1})} \boldsymbol{v}_\perp
	\left[\frac{e^{i(\omega - \boldsymbol{k} \cdot \boldsymbol{v}) \delta t} -
	1}{i(\omega - \boldsymbol{k} \cdot \boldsymbol{v})}\right] .
\end{equation}
Here the wave vector $\boldsymbol{k}$ points in the observation direction 
(to the observer) and has magnitude ${k} = \omega n/c$, with $c/n$ the speed of light in the medium and $n$ its refractive index. We will use throughout this paper a refractive index characterized by the refractivity ${\cal R}=n-1 = a \rho$ proportional to the air density $\rho$ with $a=0.25~\mathrm{g^{-1}\,cm^3}$. In magnetic media $\mu_r$ is the relative permeability of the medium.
Here, ${\boldsymbol x_1}$ is the vector position of the start of the track, and ${\boldsymbol R}$ is the vector position of the observer relative to some arbitrary origin such as the start point of the shower with $R=\vert{\boldsymbol R}\vert$ assumed to be large enough so that the contribution of the track to the electric field can be obtained in the Fraunhofer limit. 
In Eq.\,(\ref{eq:ZHS}), ${\boldsymbol{v}}_\perp = - \hat{ 
\boldsymbol{k}} \times (\hat{\boldsymbol{k}} \times \boldsymbol{v})$
is the projection of $\boldsymbol{v}$ perpendicular to the unit vector $\hat{\boldsymbol{k}} = \boldsymbol{k}/|\boldsymbol{k}|$ (pointing to the observer) in the $\boldsymbol{v}\boldsymbol{k}$ plane (containing the track and the observer). 
More recently, an equivalent formulation has been developed for the time-domain which has been implemented in the ZHS code~\cite{Alvarez-Muniz:2010wjm}. The approach to calculate the radio emission as a coherent sum of contributions from sub-tracks using Eq.\,(\ref{eq:ZHS}), or the equivalent one in the time domain, is also referred to as the "ZHS algorithm". This is exactly the approach implemented in the AIRES Monte Carlo simulation of air showers \cite{Sciutto:1999jh,AIRES_19_04_00} to give ZHAireS~\cite{AlvarezMuniz2012}, and a very similar algorithm~\cite{James:2010vm} is used in CORSIKA \cite{CORSIKA} to give CoREAS~\cite{Huege:2013vt}, 
the two leading Monte Carlo programs for simulation of radio emission from atmospheric showers. 

The positions and times of the extremes of these tracks are readily available by Monte Carlo design, and they can be used to compute the components of the electric field in the frequency and time domains. The electric field at the observer's location is given by superposition of contributions from each individual track, taking into account the relative time delays or phase shifts between different tracks because of their different positions and times. The calculation was originally made in the Fraunhofer limit with respect to the whole shower neglecting magnetic effects. In this case the field scales with distance to the shower and, due to the cylindrical symmetry, it is only necessary to give the observing angle relative to the shower axis, or "off-axis" angle. Later, the ZHS code was extended to the Fresnel region what implies further track subdivisions to obtain smaller particle propagation steps, so that the tracks themselves are in the Fraunhofer region relative to the observer~\cite{Garcia-Fernandez:2012urf}. In this case, full account of the time delays to the observer have to be made to add the contributions from the different tracks. Calculations at a given observer position require the absolute distance relative to each shower emission point in addition to the off-axis angle.  

Equation~(\ref{eq:ZHS}) contains important properties of the pulses. The term in brackets simply becomes $\delta t$ when the phase $(\omega-{\boldsymbol k} \cdot {\boldsymbol v}) \delta t$ is small. When multiplied by ${\boldsymbol v_\perp}$ gives the well-known result that the pulse amplitude is proportional to the tracklength projected onto the perpendicular to the observer direction in the $\boldsymbol{v}\boldsymbol{k}$ plane, ${\boldsymbol v_\perp} \delta t$~\cite{zas1992electromagnetic}. This limit applies both to sufficiently low frequencies and also to observation sufficiently close to the Cherenkov angle, given by the condition $\omega-{\boldsymbol k} \cdot {\boldsymbol v}=0$. 
The relative phase between different particles (or tracks) is given by the phase factor $(\omega  t_1 -{\boldsymbol k} \cdot {\boldsymbol x_1})$, where $t_1$ is the start time of the track. In the low frequency limit this phase is small and all particles contribute coherently. The spectral amplitude is proportional to the frequency, as can be directly read from the amplitude of the individual tracks in Eq.~(\ref{eq:ZHS}). This phase factor can be rewritten in cylindrical coordinates with longitudinal axis along the $z$-axis (assumed to be the shower axis without loss of generality) and $(r,\phi)$ defining the position in the $\boldsymbol{\hat{x}}\boldsymbol{\hat{y}}$ plane. In the simplified case of the shower front moving as a plane front at the speed of light in the $z$ axis, $t_1$ could be simply replaced by $z/c$. Clearly, for a realistic shower, $t_1$ will be larger because the particles accumulate delays with respect to this plane due to trajectories deviating from the $z$ direction and also due to particles not moving at the speed of light~\cite{cazon2004time}. We will denote these time delays as $t_D$; they characterize the curvature and width of the shower front and play an important role in air showers~\cite{alvarez2012coherent}. We can thus write $t_1=z/c+t_D$ to obtain: 
\begin{equation}
R\vec{E}(\omega, {\boldsymbol x})=\frac{e \mu_r i \omega}{2 \pi \epsilon_0 c^2} ~ e^{i k R} 
~{\boldsymbol v}_{\perp}\delta t ~ 
\exp{iw \left[\frac{(1-n\cos\theta)z}{c} + t_D - \frac{n r \,\sin\theta \cos\phi}{c}\right]}.
\label{eq:zhs_formula_cher}
\end{equation}

The relevant time delays for the relative phase factor in brackets in Eq.~(\ref{eq:zhs_formula_cher}) are separated into three terms. The first one cancels exactly for observers in the Cherenkov direction, defined by $n \beta \cos\theta_C =1$ with $\beta=v/c$. For an observer in the Fraunhofer limit at $\theta_C$ the $z$ coordinate is irrelevant for interference. As the observer position moves away from the Cherenkov angle this term is responsible for interference between early and late parts of the longitudinal shower development along $z$. The second term, $t_D$, corresponds to the time delays of the particles relative to the plane moving at the speed of light in the $z$ direction. The third one, proportional to $r$, gives the relative delay of the emission from a given particle track which is due to the lateral displacement (perpendicular to shower axis) of its start point $r$ as seen from the observer's position. 
Separating the time delays in this manner allows a more intuitive explanation of the behavior with frequency of the Fourier spectrum of the electric field amplitude obtained in air under different circumstances. Particle tracks with small phases corresponding to the three terms in Eq.\,(\ref{eq:zhs_formula_cher}) contribute coherently and give the bulk of the emission. For sufficiently low $\omega$, typically below the inverse of the term in brackets in Eq.\,(\ref{eq:zhs_formula_cher}), all particles in the shower contribute coherently. However, as the frequency rises the number of particles emitting coherently gets reduced, and particles with phases larger than about $\pi/2$ can be practically ignored because their total contribution becomes negligible relative to those that emit coherently.

\subsection{The ZHS code with magnetic deflections}

Realistic electromagnetic pulses emitted in showers developing in a low density media such as air cannot be obtained without considering the bending of the particle trajectories in the magnetic field of the Earth. The time delays associated to magnetic deflections are proportional to the distances involved and hence scale with the inverse of the density so that they play a very important role in the coherence properties of the radio pulses in air. As a result, they can also be expected to be more important for showers developing higher in the atmosphere, where the atmosphere is less dense, as it is the case for very inclined cosmic-ray showers. 

A new version of ZHS has been developed to account for magnetic deflections. It was developed trying to keep an accurate description of the time delays of the particles by carefully considering the effects of energy loss and multiple elastic scattering combined with those induced by magnetic deflections. The algorithm developed for this purpose and implemented in the original ZHS is described in detail in Appendix~\ref{Appendix1}. Shower development and radio emission in air obtained with the upgraded version of ZHS accounting for magnetic deflections, are compared to the results obtained with AIRES and ZHAireS in constant density air as a means to test and validate the performance of ZHS.

\subsection{Comparison of ZHS with AIRES and ZHAireS Monte Carlo simulations}
\subsubsection{Comparison of longitudinal and lateral profiles}
Firstly, shower development in air, as obtained with the modified version of ZHS accounting for magnetic effects, has been compared to that obtained with AIRES\,\footnote{AIRES version 19.04.08 was used for all comparisons \cite{AIRES_19_04_00}.} in constant density air. Just for these comparisons the kinetic energy threshold of electrons and positrons and total energy threshold of photons in ZHS, have been set to 1 MeV to match the default threshold for knock-on interactions used in AIRES. This energy is also used to separate continuous energy loss for electrons and positrons from discrete interactions. Additionally, the M\o ller cross section used in AIRES has been modified at energies in the few MeV range to better match that of ZHS, and the implementation of continuous energy loss in AIRES has been replaced with the approach used in  ZHS~\cite{zas1992electromagnetic}. These two changes were made to minimize the effects that different implementations of these two processes can have on shower development and that could mask differences attributable to the implementation of magnetic deflection\cite{Alvarez-Muniz:2010hbb}.

\begin{figure}[ht]
	\centering
	\subfigure{\includegraphics[width=0.49\linewidth]{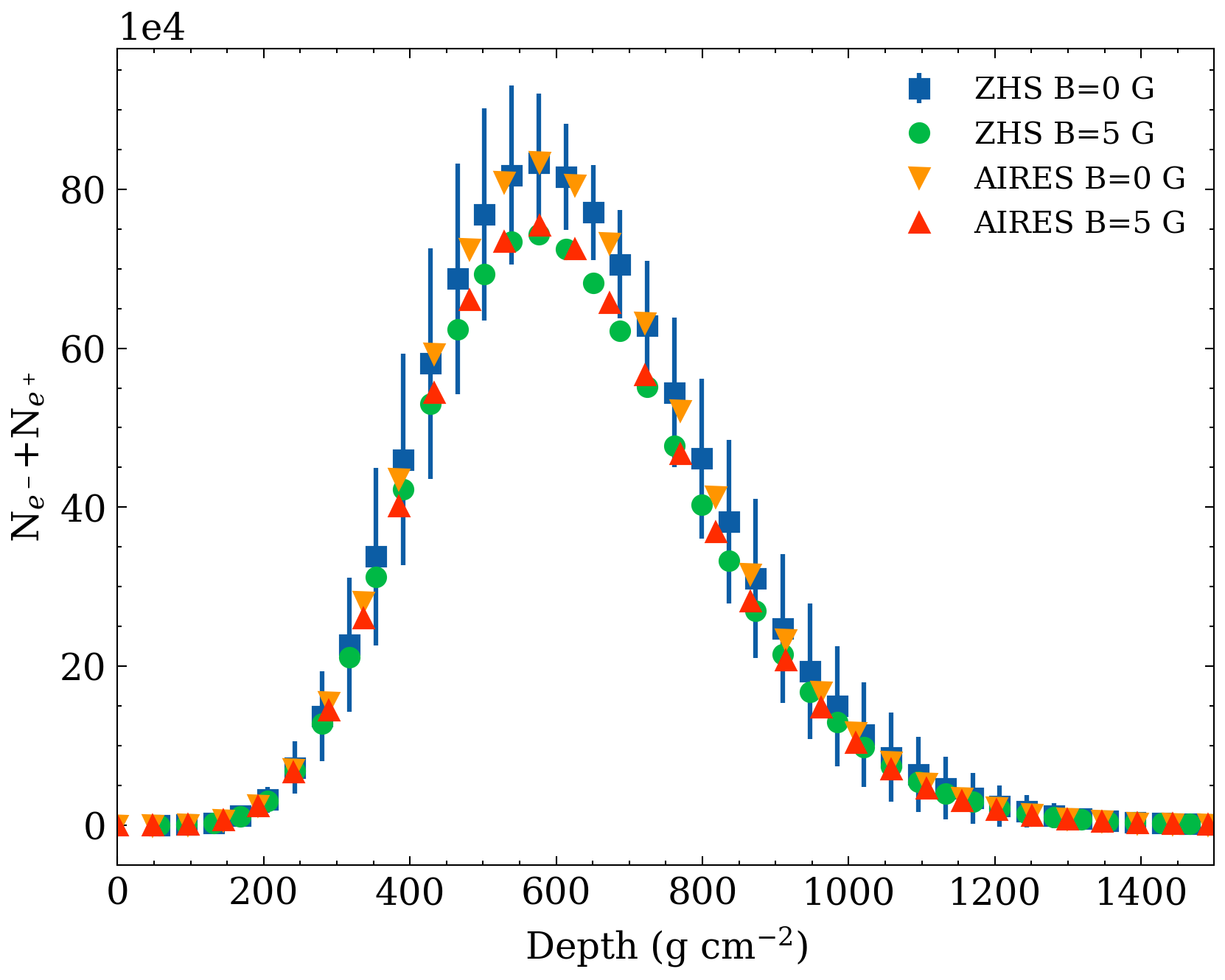}}
	\subfigure{\includegraphics[width=0.49\linewidth]{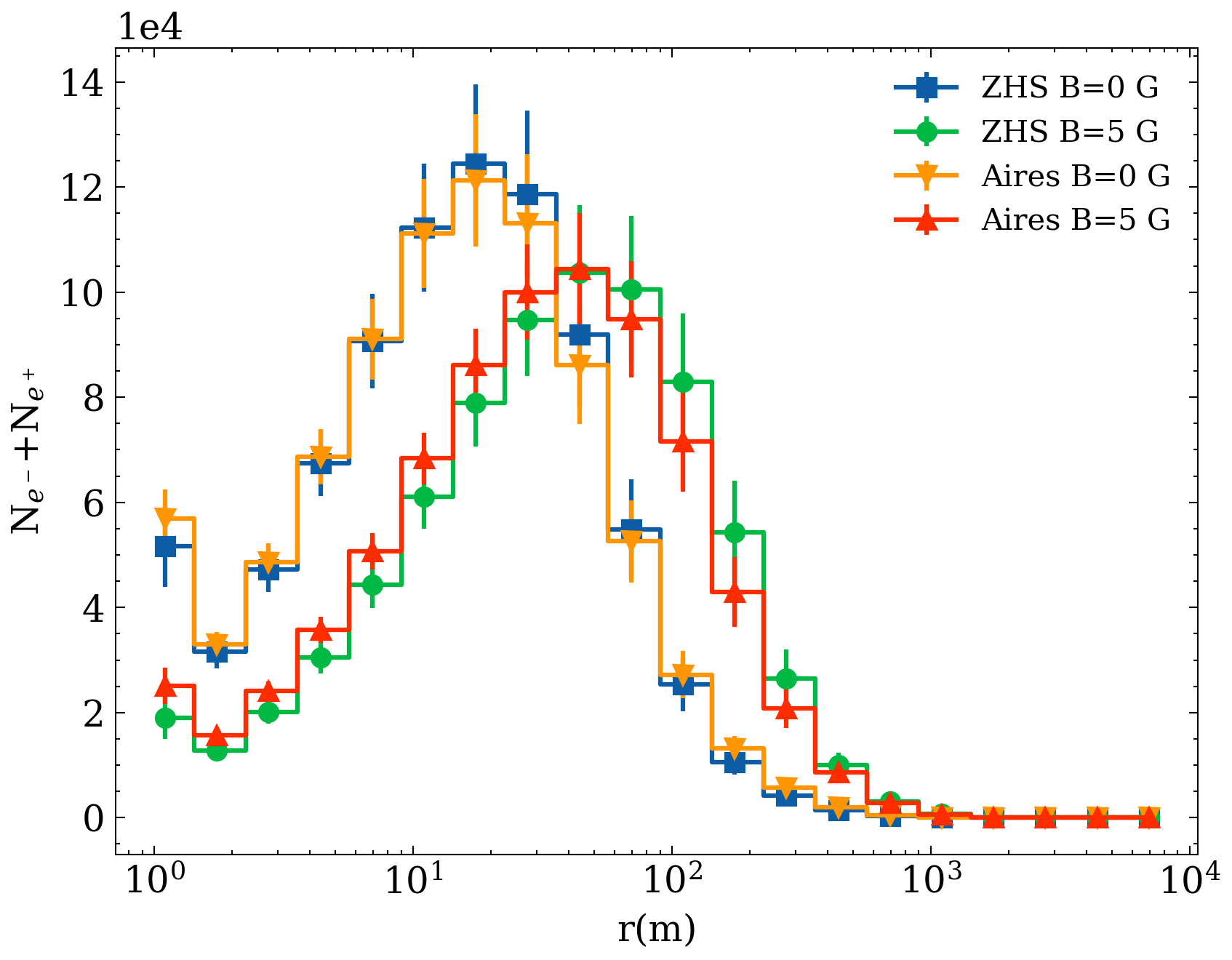}}
	\caption{\label{fig:aires_comparison} Average of the longitudinal (left) and lateral (right) profiles of 200 electron-induced showers of 1 PeV energy with a  very high intensity magnetic field ($B=5\,\mathrm{G}$) perpendicular to shower axis, and without the magnetic field. The error bars indicate the standard deviation. The longitudinal profiles shown only contain the standard deviations for ZHS with $B=0\,{\rm G}$ to improve readability.
 }
\end{figure}

We have simulated two sets of 200 electron-induced showers of 1\,PeV energy first without magnetic field for a constant density of air of $1.2~10^{-3}~\mathrm{g\,cm}^{-3}$ (matching sea level conditions) with ZHS and ZHAireS. Similar tests without magnetic field and for electromagnetic and hadronic showers have been done in the past in ice~\cite{Alvarez-Muniz:2010hbb}. Comparisons of the average longitudinal development of electrons and positrons in the shower are shown in the left panel of Fig.~\ref{fig:aires_comparison}, and comparisons of the average lateral distributions of electrons and positrons in a plane perpendicular to shower axis at a depth close to shower maximum are shown in the right panel of Fig.~\ref{fig:aires_comparison}. The shape of the longitudinal profiles agrees at a level of less than $4\%$\footnote{A small constant shift of order $20~\mathrm{g\,cm}^{-2}$ is required to have the profiles aligned at the same depth, but such a shift is irrelevant for the radio pulses in the far distance limit.}. The average lateral distributions at shower maximum also display a good agreement at a level below $7\%$. 

The simulations have been repeated with a magnetic field of intensity $B=5\,\mathrm{G}$, about 10 times higher than characteristic values of the Earth's magnetic field, to significantly modify the lateral distribution of the shower. The results are also shown in Fig.~\ref{fig:aires_comparison}. The longitudinal development of electrons and positrons predicted in both ZHS and ZHAireS with magnetic field deflections agree at the level of $2\%$. When comparing simulations with and without magnetic field, it can be noted that the number of particles at shower maximum decreases by $\sim 10\%$ when the magnetic field $B=5\,\mathrm{G}$ is turned on in the simulations. This is due to the deviations of particles from the shower axis so that their trajectories advance smaller distances in the longitudinal direction and the number of particles crossing planes normal to the shower axis gets reduced. The agreement between the shapes of the lateral distributions at the depth of shower maximum predicted with the two simulations is below $14\%$. As the magnetic field is turned on, the lateral distribution becomes wider as could be expected due to the magnetic deflections. This can be quantified calculating the distance from the shower axis containing $90\%$ of the electrons and positrons that increases from $\sim 60$ m when there is no magnetic field to $\sim 140$ m with a magnetic field intensity of $B=5\,\mathrm{G}$. These tests at the level of shower profiles and lateral distributions shown in Fig.~\ref{fig:aires_comparison} give further confidence on the correct implementation in the ZHS program of deviations of charged particle tracks due to the Lorentz force. 

\subsubsection{Comparison of radio pulses}

The induced pulses obtained with the upgraded version of ZHS have also been compared to those obtained with ZHAireS\footnote{ZHAireS version 1.0.30 was used for all comparisons.}. We have compared pulses from individual showers rather than average pulses from many showers. This procedure requires comparing different selected showers with a similar longitudinal development in order to isolate the effects on the pulses of the implementation of geomagnetic deviations from those induced by the differences in shower development. Inevitably the pulses will however display differences associated to the different longitudinal and lateral distributions of the individual showers. 

\begin{figure}[ht]
\centering
\includegraphics[width=1.0\linewidth]{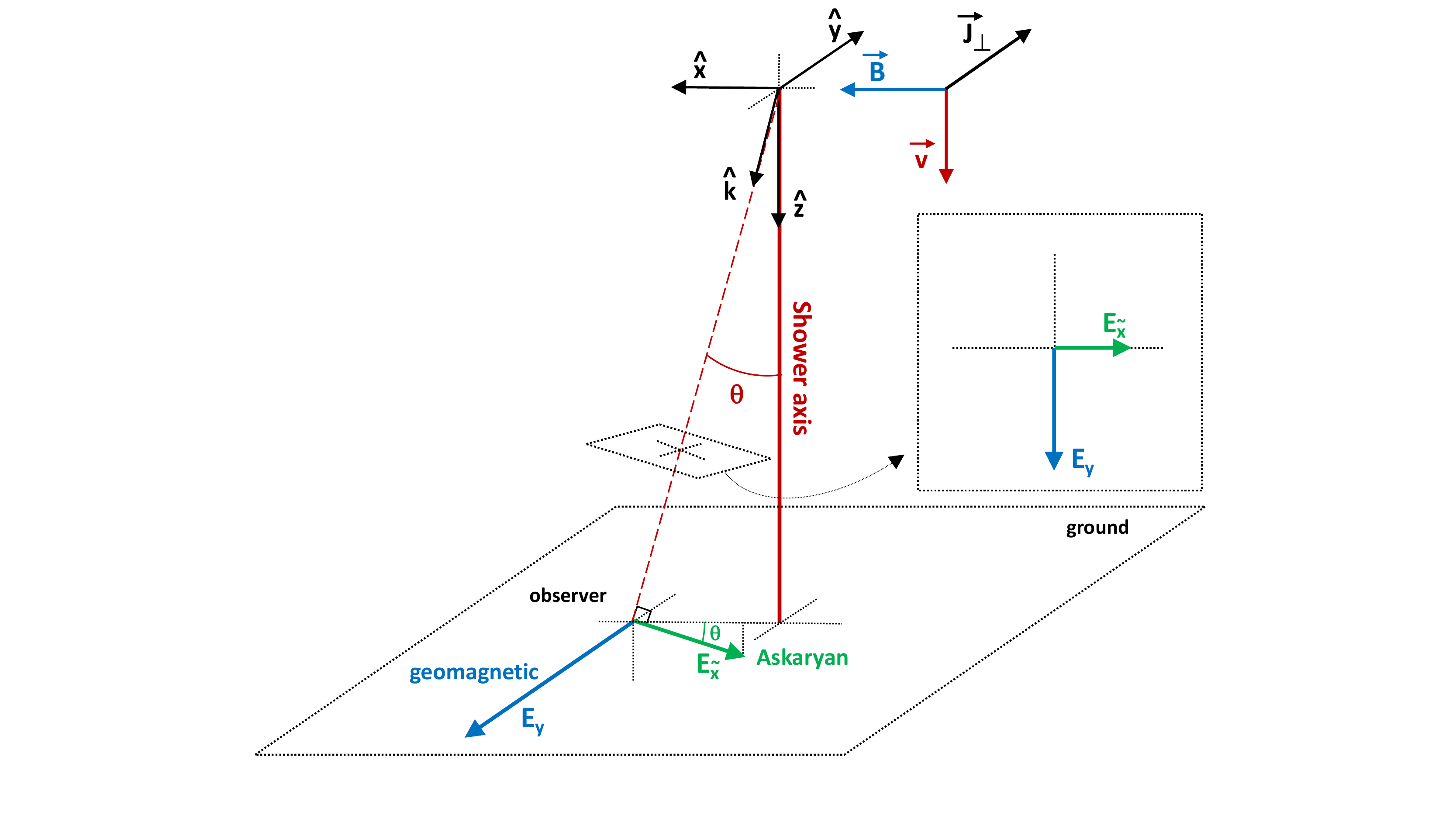}
\caption{Sketch of the geometry adopted in this work. The shower axis is parallel to ${\boldsymbol{\hat{z}}}$, the observer position is on the ${\boldsymbol{\hat{x}}} {\boldsymbol{\hat{z}}}$ plane, and the magnetic field is parallel to ${\boldsymbol{\hat{x}}}$. The azimuthal angle of the observer position is $\phi=0$, defined with respect to the $x$ axis. In this particular geometry, the $E_y$ component of the electric field (in blue) is due to the geomagnetic effect, and the orthogonal component $E_{\tilde{x}}$ (almost parallel to the ${\boldsymbol{\hat{x}}}$ axis) is due to the Askaryan effect (in green).
}
\label{fig:Geometry}
\end{figure}

Throughout this paper we use a fixed geometry for the shower axis (along $\boldsymbol{\hat{z}}$ in Fig.~\ref{fig:Geometry}) and the magnetic field which is assumed to be orthogonal to the shower axis (along $\boldsymbol{\hat{x}}$ in Fig.~\ref{fig:Geometry}) to maximize the effect of the magnetic field on the radio emission. The observer is located at the Cherenkov angle so that the coherence of the emission is expected to be largest. In this case, the transverse current $\boldsymbol{J_\perp}$, is parallel to $\boldsymbol{\hat{z}} \cross \boldsymbol{\hat{x}}$, pointing in the positive $\boldsymbol{\hat{y}}$ direction, and the observer is assumed to lie in the $\boldsymbol{\hat{x}}\boldsymbol{\hat{z}}$ plane, corresponding to an azimuth angle $\phi=0$. The electric field induced by the transverse current, due to the geomagnetic effect, is polarized along the ${\boldsymbol{\hat{y}}}$ axis. On the other hand, the component of the current along the ${\boldsymbol{\hat {z}}}$ axis is due to the excess charge that develops in the shower, the Askaryan effect. In this geometry, the electric field due to the Askaryan effect is perpendicular to $\boldsymbol{k}$ and contained in the $\boldsymbol{\hat{z}}\boldsymbol{\hat{x}}$ plane, and is orthogonal to the polarization due to the transverse geomagnetic current. As the observer is located at an off-axis angle equal to the Cherenkov angle in the $\boldsymbol{\hat{z}}\boldsymbol{\hat{x}}$ plane, which is typically of order one degree in air, this component is practically (but not completely) parallel to the $x$-axis. We will refer to this polarization as the $\tilde{x}$ polarization to avoid this confusion. By choosing this geometry the geomagnetic effect and the excess charge or Askaryan effect are separated in two quasi-orthogonal polarizations, allowing us to study each of these mechanisms independently. Such geometry will be kept for all the simulations that will be described in the following sections. 

\begin{figure}[ht]
\centering
\subfigure{\includegraphics[width=0.49\linewidth]{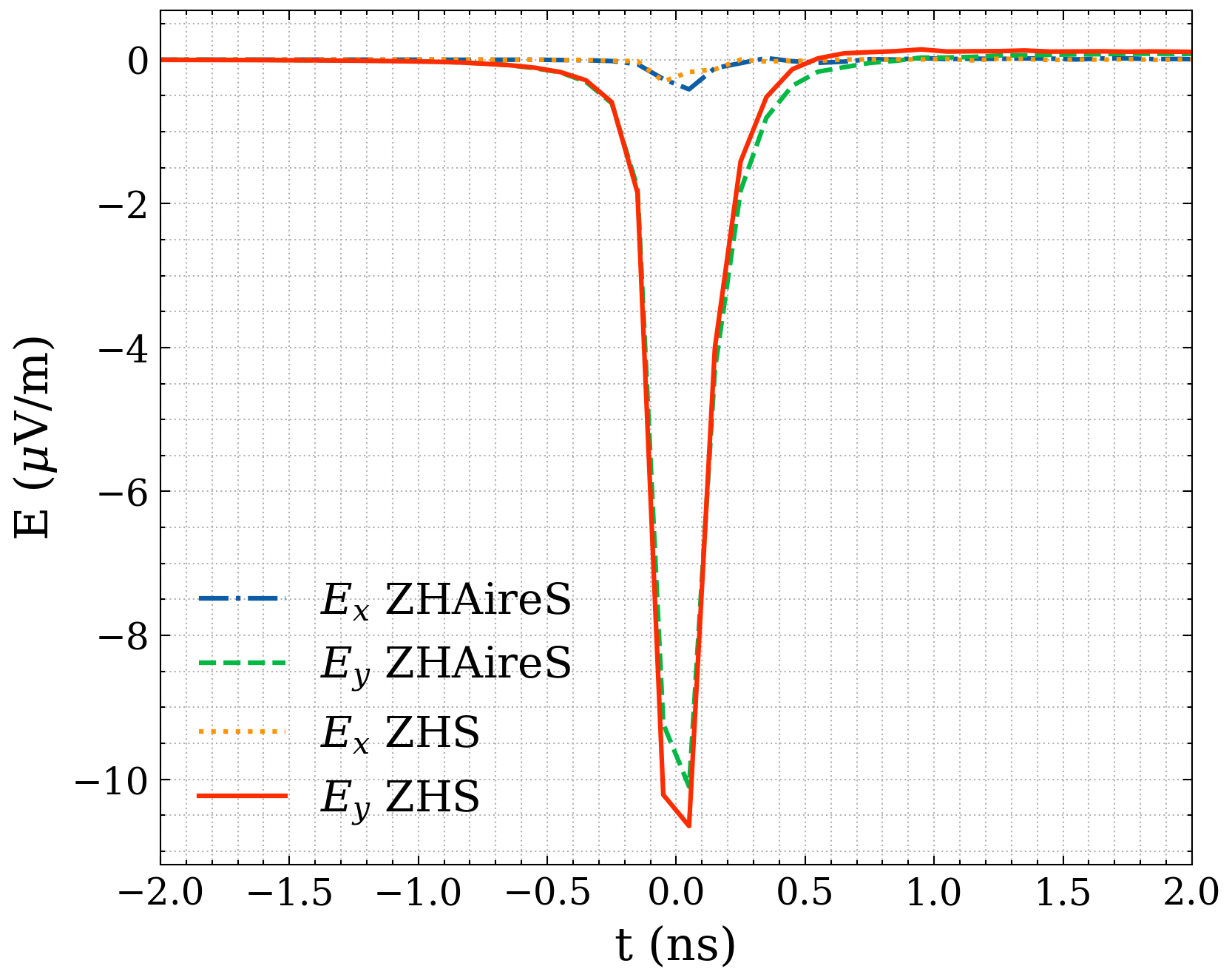}}
\subfigure{\includegraphics[width=0.49\linewidth]{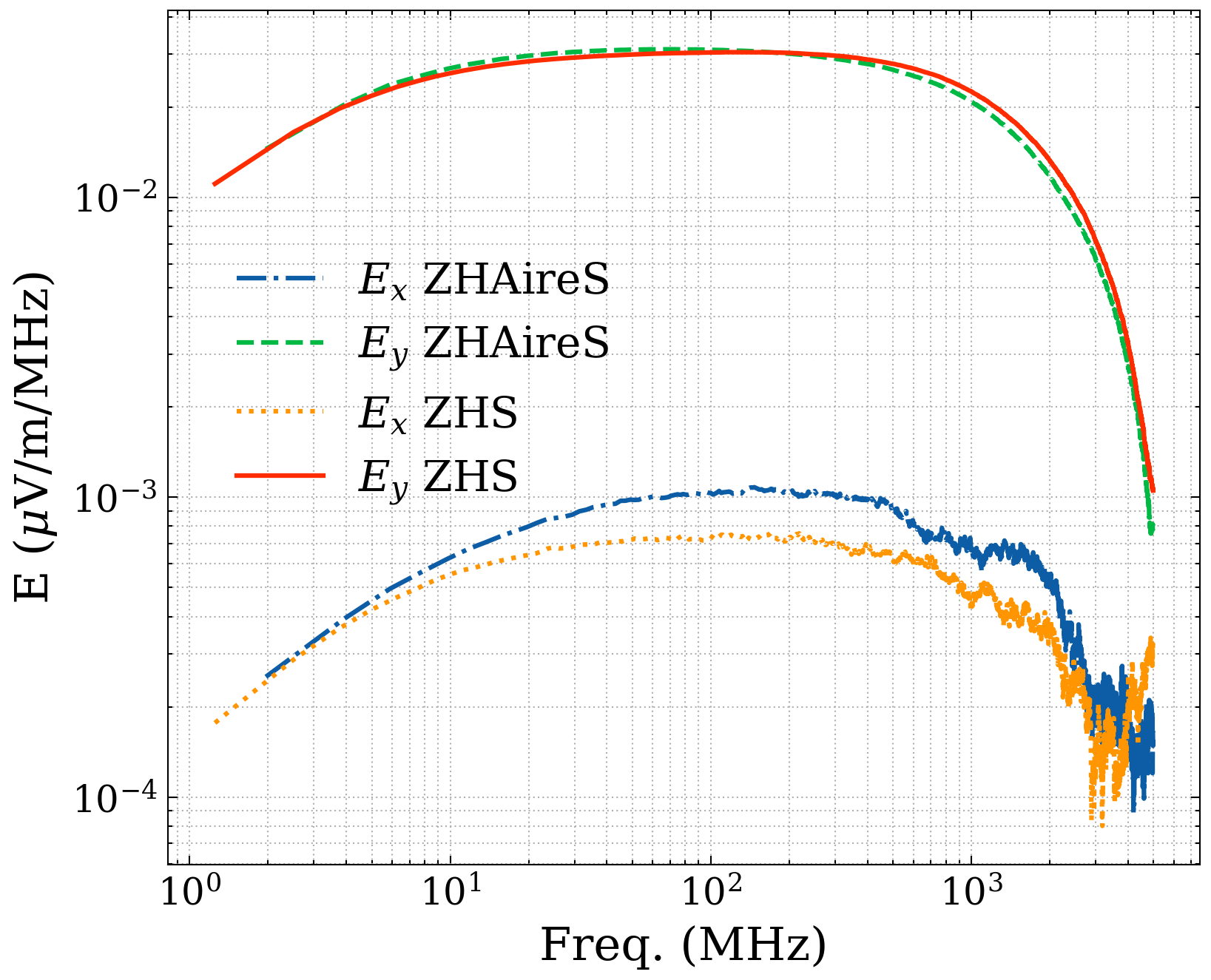}}
\caption{Comparison of the $E_{\tilde{x}}$ and $E_y$ components of the electric field predicted with ZHAireS and the upgraded ZHS accounting for magnetic deflections (see geometry in Fig.\,\ref{fig:Geometry}). The components are shown in the time (left) and frequency domain (right). The simulated showers are initiated by a 1 PeV electron with a magnetic field of intensity $B=5\,{\mathrm G}$ pointing along the $x-$axis in Fig.\,\ref{fig:Geometry}. The observer is located at coordinates $(x,y,z)=(2.444, 0, 100)$ km relative to the shower injection point, corresponding to observation at approximately the Cherenkov angle $\theta_C=1.4^\circ$.}
\label{fig:pulse_comparison}
\end{figure}

Adopting this geometry, the comparison of the pulses predicted by ZHS and ZHAireS is made for an observer position in the $\boldsymbol{\hat{x}}\boldsymbol{\hat{z}}$ plane, viewing the shower from the Cherenkov direction, with an off-axis angle $\theta=1.4^\circ$, where the coherence of the pulses is expected to be maximum and the pulses can be expected to be shortest. The $\tilde{x}$ and $y$ components of the electric field obtained with ZHS and ZHAireS are shown in Fig.\,\ref{fig:pulse_comparison} both in the time (left panel) and frequency (right panel) domains. The plot illustrates the level of agreement between the calculations. The comparison in the frequency domain shows agreement for the geomagnetic component at a similar level of less than $5\%$ below 1 GHz. The amplitudes for the Askaryan component with the ZHS program are of order $30\%$ below those of ZHAireS. Part of this effect is because different showers are being compared. We also note that despite the magnetic field intensity being unrealistically high in this case, the Askaryan component is typically not expected to have a large impact in practice in many scenarios. The pulse is the result of the collective interference of all the charged particle tracks and is affected by differences in the space-time distribution of the charges in the shower. The comparison of the pulses serves to test the relative differences between the space and time coordinates of the charged tracks in both simulations, that agree at the level of the inverse frequency of the pulse, which in this case is of order 30 cm and 1 ns, corresponding to frequencies of order 1 GHz. 

\subsection{Spectral amplitude of radio pulses in air with ZHS} 
\label{AskaryanInAir}

We show the results of the simulations of the radio pulses in air in the frequency domain and in the Fraunhofer limit. As could be expected and will be shown, the study of the frequency spectrum is more adequate to relate the characteristics of the pulses to the shower properties 
and to deduce scaling properties. By choosing an observer in the far distance limit and in the Cherenkov direction we maximize the coherence of the emission, and we eliminate the effect of observation distance so that the position of the observer in spherical coordinates only needs two angles, the off-axis angle which is the polar angle relative to the shower axis (chosen to be the $z$ axis), and the azimuth angle, $\phi$, measured in the $\boldsymbol{\hat{x}}\boldsymbol{\hat{y}}$ plane as shown in Fig.\,\ref{fig:Geometry}. The pulse will still depend on the orientation of the magnetic field of the Earth relative to the shower direction, the magnetic field intensity and the density of air, and on the observation position relative to the shower. Reducing the number of explored variables on which the radio pulse depends is quite important to find the underlying scaling behavior. 

\begin{figure}[hbt]
\centering
\subfigure{\includegraphics[width=0.49\linewidth]{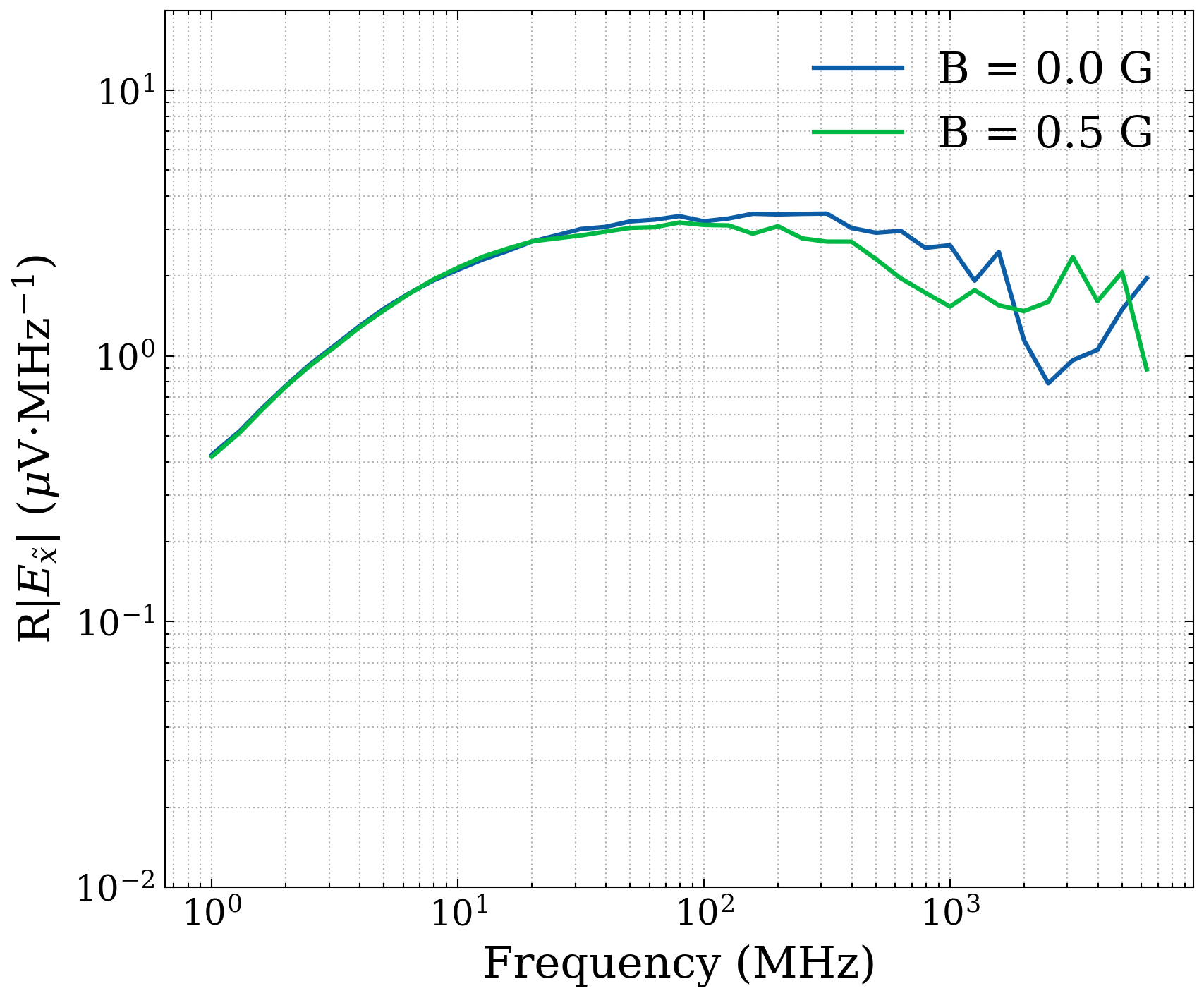}}
\subfigure{\includegraphics[width=0.49\linewidth]{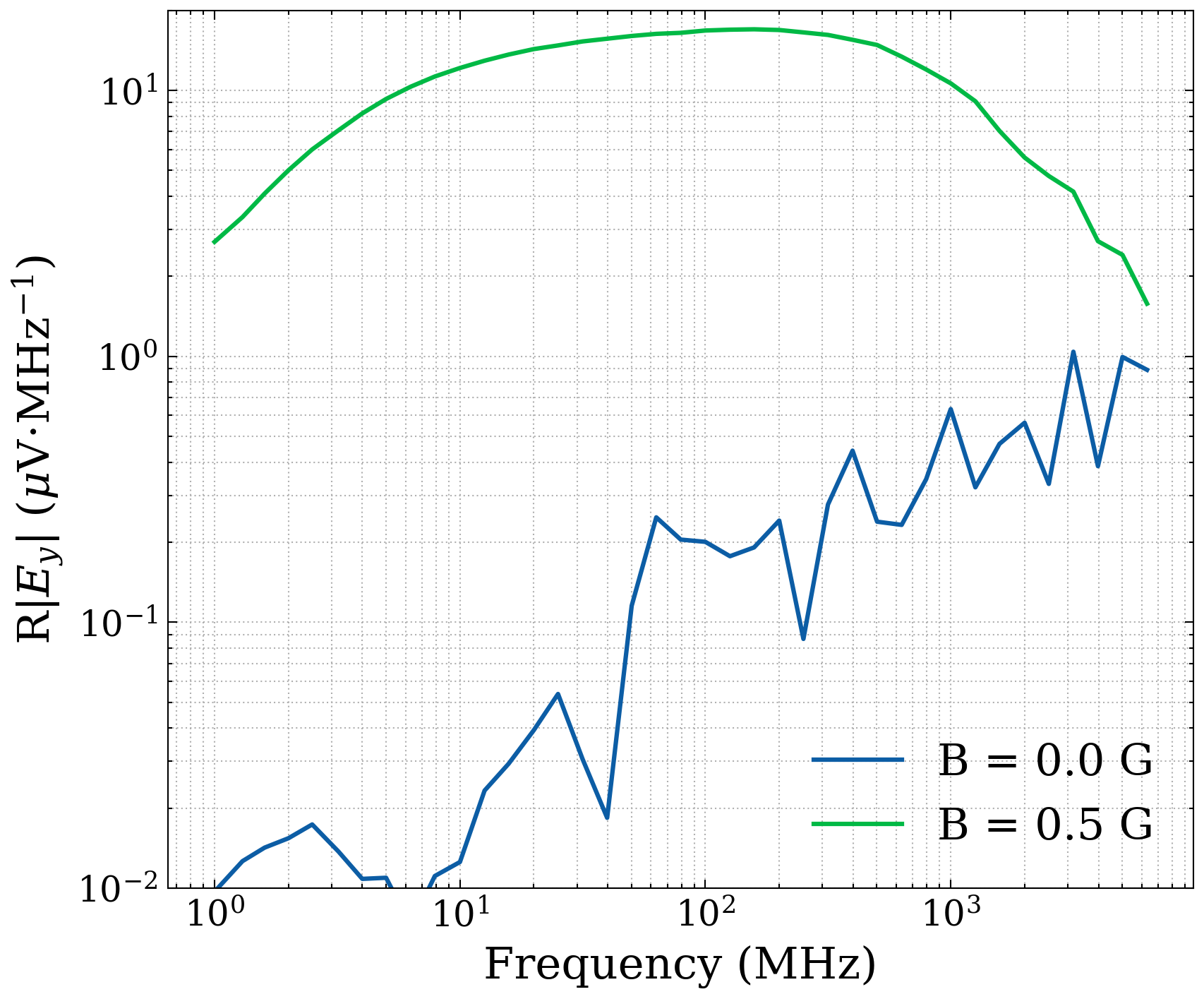}}
\caption{Frequency spectrum of the radiopulse for an observer at the Cherenkov angle in the Fraunhofer regime, induced by a 100 TeV electron shower in constant density air with $\rho=0.0012~{\rm g\,cm^{-2}}$. The left panel shows the $E_{\tilde{x}}$ polarization of the electric field, which in the geometry sketched in Fig.\,\ref{fig:Geometry} is mainly due to the excess charge in the shower. The right panel shows the $E_y$ polarization dominated by the geomagnetic emission when the magnetic field intensity is perpendicular to shower axis (see Fig.\,\ref{fig:Geometry}) and set to $B=0.5\,\mathrm{G}$.} 
\label{fig:ice_radiopulse} 
\end{figure}

We have obtained radio pulses in showers initiated by 100 TeV electrons in air of constant density $\rho=0.0012~\mathrm{g\,cm^{-3}}$ both with a magnetic field intensity of $B=0.5\,\mathrm{G}$ and $B=0$ (no magnetic field). The pulses have been obtained for an observer in the Fraunhofer limit with an off-axis angle equal to the Cherenkov angle for the geometry sketched in Fig.~\ref{fig:Geometry}. The results for the $E_x$ and $E_y$ polarizations of the electric field are shown separately in the two panels of Fig.~\ref{fig:ice_radiopulse}, comparing the pulses obtained with and without the magnetic field. The $E_x$ component is similar in both cases since this is mainly due to the excess charge developed along the longitudinal development of the shower which is largely independent of the magnetic field intensity. In contrast, the $E_y$ component, mostly associated to the geomagnetic effect, is very much enhanced for the magnetic field intensity $B=0.5\,\mathrm{G}$. Even when $B=0$ there is a remaining $E_y$ component that can be associated to the transverse velocity of electrons and positrons which can be expected to have have a random distribution. The contributions from all particles are thus incoherent because they have random polarizations and the electric field amplitude only scales with the square root of the frequency\footnote{Note the obtained amplitudes display large fluctuations relative to the expected $\omega^{1/2}$ behavior because showers of energy 100 TeV have a relatively small number of particles $\sim 10^5$.} as can be inferred from the right panel of Fig.~\ref{fig:ice_radiopulse}.

\begin{figure}[ht]
\centering
\subfigure{\includegraphics[width=0.8\linewidth]{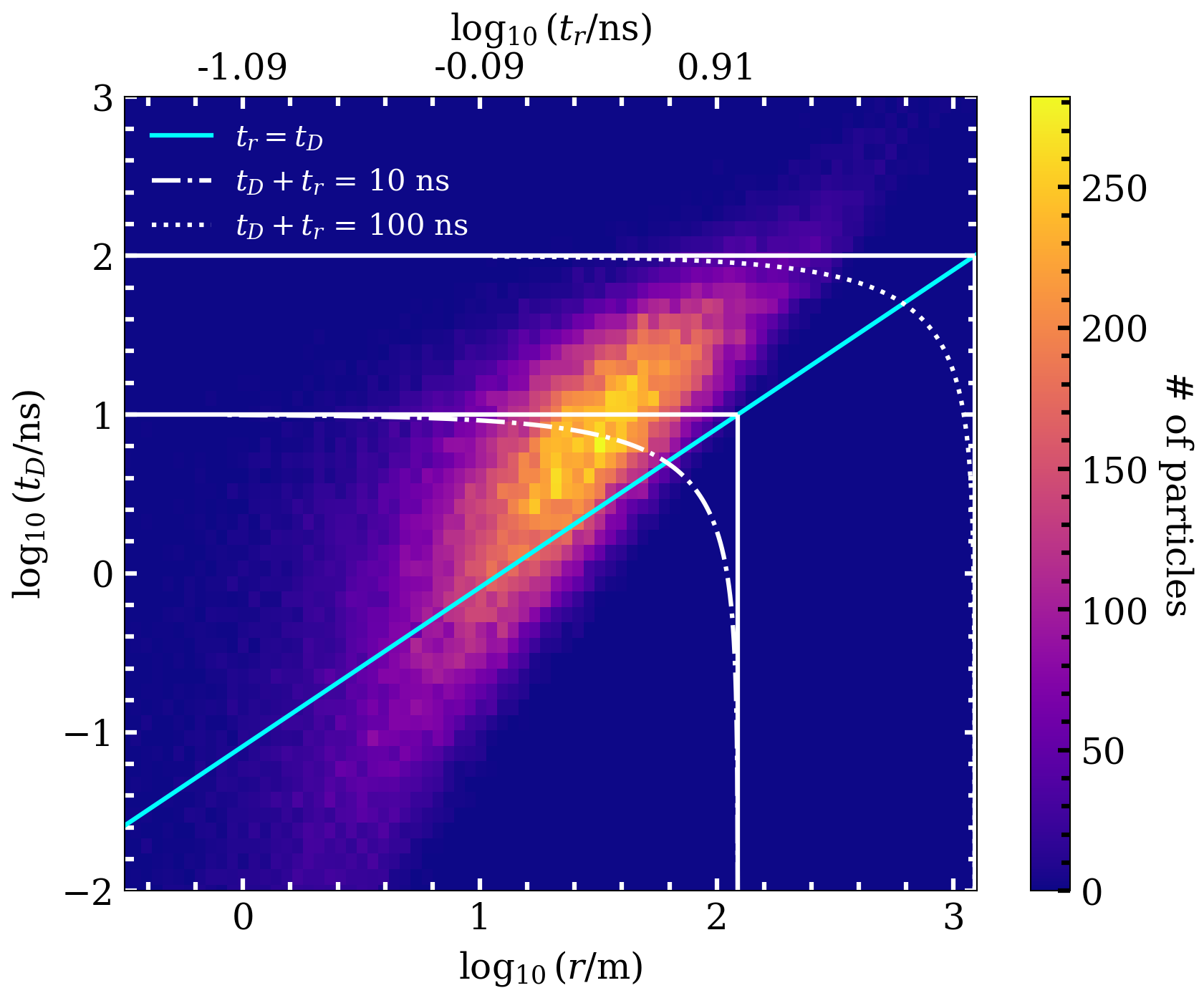}}
\caption{Distribution of particles in two-dimensional space given by $r$, the lateral distance, and $t_D$, the time delay with respect to a plane moving at the speed of light and perpendicular to the shower axis. The distributions shown correspond to 100 TeV electron showers in air with constant density $\rho=0.0012~{\rm g\,cm^{-2}}$, for a magnetic field perpendicular to shower axis and of intensity $B=0.5\,\mathrm{G}$. The plots correspond to the particles that cross a plane perpendicular to the shower axis, located 15 radiation lengths after injection, close to the maximum of the shower. The top scale for the $x$ axis of the figure corresponds to the maximal delay at a distance $r$ due to the lateral displacement of the particles, given by $t_r^{max}-\simeq nr\sin \theta/c$ (see text). For an observer in the Cherenkov direction $t_D$ and $t_r$ are the only relevant delays (see text and Eq.~(\ref{eq:zhs_formula_cher})). The light blue line corresponds to $t_D=t_r^{\rm max}$ and delimits a region above it in which $t_D$ dominates. Below this line, the lateral delay $t_r$ can become more relevant (for $\cos \phi=\pm 1$). The dash-dotted and dotted white lines correspond to $t_D+t_r^{\rm max}=10$ and $100$~ns respectively. The particles in the region to the bottom left of these two lines contribute coherently (see text). 
}
\label{fig:time_delay_hist} 
\end{figure}

The frequency spectrum of the radiated electric field has three different regimes as can be clearly appreciated in Figs.~\ref{fig:pulse_comparison} and \ref{fig:ice_radiopulse}. At the lowest frequencies, the spectral amplitude scales with $\omega$. However, deviations from such linear behavior arise at frequencies above about 5 MHz. The spectrum flattens in a region between 20 and 200 MHz, and above 200 MHz the amplitude drops as the frequency rises. These features can be explained by considering the number of particles that contribute coherently. At low enough frequencies all particles contribute coherently with roughly the same phases in Eq.~(\ref{eq:zhs_formula_cher}) and the spectrum naturally scales with $\omega$ as expected from Eq.~(\ref{eq:zhs_formula_cher}) which explicitly displays this behavior. The gradual flattening that starts at 5 MHz can be associated to an increasing number of particles that loose coherence, mainly due to their delays $t_D$ with respect to a plane front moving at the speed of light, or equivalently to the curvature and width of the shower front~\cite{alvarez2012coherent,Alvarez-Muniz:2011wcg}. The delays $t_D$ can be appreciated in Fig.~\ref{fig:time_delay_hist} where the distribution in $(r,t_D)$ space, with $r$ the distance to the shower axis, is shown for electrons and positrons that cross a plane perpendicular to the shower direction near the depth of shower maximum. As a rough approximation, at frequencies below 5 MHz the particles with delays exceeding 200 ns can be assumed not to contribute coherently to the field. This occurs for only a small fraction of the particles, those lying above a horizontal line at $t_D=200$~ns ($\log_{10}(t_D/\mathrm{ns})\simeq 2.3$), causing slight deviations from the linear behavior of the amplitude with $\omega$. However, when the frequency increases further between 20 and 200 MHz, corresponding to delays between $t_D=50\,\mathrm{ns}$ ($\log_{10}(t_D/\mathrm{ns})\simeq 1.7$) and $t_D=5\,\mathrm{ns}$ ($\log_{10}(t_D/\mathrm{ns})\simeq 0.7$), corresponding to the bulk of the distribution in Fig.~\ref{fig:time_delay_hist}, the amount of particles that contribute coherently decreases nearly linearly with $\omega$, compensating the linear scaling with $\omega$ in Eq.\,(\ref{eq:ZHS}) flattening the spectrum in this region. 

The $r$-scale of the $x$ axis in the histogram of Fig.~\ref{fig:time_delay_hist} can be translated into a maximum time delay $t_r$ due to the lateral distribution of the particles given by $t_r^{\rm max}=n r \sin \theta/c$ (obtained setting $\cos \phi = \pm 1$ in the third term of the phase factor in Eq.\,(\ref{eq:zhs_formula_cher})). The conversion of $r$ to $t_r^{\rm max}$ is shown in the scale at the top of the histogram in Fig.\,\ref{fig:time_delay_hist}. The light blue line corresponds to $t_D=t_r^{\rm max}$, above it the particle delays $t_D$ dominate, and below it the maximal delay associated to the lateral position of the particles $t_r^{\rm max}$ (for $\phi=0^\circ$ and $180^\circ$) is larger. We note that $t_r$ is below $\sim 50$~ns for the bulk of shower particles, so lateral delays are thus irrelevant to explain the start of the plateau in the frequency spectrum  of the electric field at $\sim 20\,\mathrm{MHz}$ in Figs.\,\ref{fig:pulse_comparison} and \ref{fig:ice_radiopulse}. However, for  frequencies above $\sim\,200\,\mathrm{MHz}$, some particles with $t_r$ exceeding $t_D \sim 5~\mathrm{ns}$ cease to contribute coherently. The plot also shows dashed-dotted and dotted lines with $t_D+t_r^{\rm max}=10$~ns, and $100$~ns respectively\footnote{The sum of $t_D$ and $t_r$ is relevant for interference.} to illustrate how, at large frequencies, some particles below constant values of $t_D$ (horizontal solid white lines) can also lose coherence because of $t_r$, more apparent in the case of $t_D=10$\,ns. For large enough frequencies, the loss of coherence is both due to particles lagging behind the shower front and spreading in the direction transverse to the shower axis. For instance, at a frequency $\nu=200\,\mathrm{MHz}$ particles above the line $t_D+t_r=1/\nu = 5$ ns in Fig.~\ref{fig:time_delay_hist} do not contribute in a fully coherent fashion. The combination of both effects makes the spectral amplitude to drop more rapidly with frequency after the flat region, as shown in Figs.~\ref{fig:pulse_comparison} and \ref{fig:ice_radiopulse}.

We note that this is a quite general description of the behavior of the frequency spectrum of radio emission in air showers, also valid in simulations with a more realistic atmosphere with density and refractive index varying with height~\cite{AlvarezMuniz2012,alvarez2012coherent}, and observed under different experimental circumstances~\cite{alvarez2012coherent,Alvarez-Muniz:2010hbb,nigl2008frequency,ANITA:2010ect,ANITA:2018sgj,Schoorlemmer:2015afa}. 

\section{Radio pulses in extensive air showers}

Coherent emission of pulses from air showers represents a rich and complex problem with many degrees of freedom that play an important role namely, the arrival direction of the shower, its orientation relative to the Earth's magnetic field and the observing direction and position. The pulses arise from two main effects (excess charge and geomagnetic deflections) that have different polarization patterns. The superposition of the two components induces an electric field that is dependent on the position of the observer, the direction of the shower axis and the orientation and intensity of the magnetic field. The transverse currents depend on both the magnetic field intensity and the density of the medium, and both have strong implications for the particle time delays. The Cherenkov angle, giving the direction where coherence is maximized, also depends on air density. 
The air density in turn changes as the shower develops and varies with the inclination of the shower. Finally, the distance to the observer in relation to the shower dimensions is also expected to have an impact on the coherence properties of the pulses.

\begin{table}[ht]
\centering
\begin{tabular}{|c|c|c|c|c|c|c|}
\hline
$\theta~(^\circ)$   & h~(km) & $\rho$~(mg~cm$^{-3}$) & $\mathcal{R} \times 10^6$ & $\theta_{C}(^\circ)$ & $D$~(km) & $\Delta\rho$~(mg~cm$^{-3}$) \\ \hline \hline
0  & 2.64 & 0.943 & 236 & 1.245 & 2.64 & 0.218 \\
30 & 3.73 & 0.845 & 206 & 1.163 & 4.31 & 0.190 \\
60 & 7.74 & 0.540 & 127 & 0.913 & 15.5 & 0.131 \\
70 & 10.2 & 0.411 & 93.5 & 0.786 & 29.7 & 0.099 \\
80 & 14.1 & 0.222 & 58.1 & 0.617 & 78.7 & 0.064 \\
85 & 17.0 & 0.141 & 40.9 & 0.519 & 170 & 0.040 \\ \hline
\end{tabular}
\caption{Characteristic parameters for cosmic ray showers of different zenith angles as labeled on the first column. The next four entries correspond to altitude, $h$, density, $\rho$, refractivity, $\mathcal{R}=n-1$ and Cherenkov angle, $\theta_{C}$ corresponding to the point at which shower maximum occurs for different zenith angles (assuming first interaction takes place at the top of the atmosphere and $X_{max}=750$\,g\,cm$^{-2}$). The next to last entry, $D$, is the distance from $X_{\rm max}$ to the impact point of the shower at ground, assumed to be at sea level altitude. The last entry, $\Delta \rho$, is the change of density that one can expect three radiation lengths before and after $X_{max}$ in shower development. The bulk of the radio emission can be assumed from particles within this interval.}
\label{tab:x_max_magnitudes}
\end{table}

As shower dimensions scale with the inverse of the density, the distance between the observer and the shower can be small or large compared to the shower dimensions depending on the type of shower, its zenith angle, and the observer position. For vertical air showers (as produced by cosmic rays) detected with antennas at ground level, the distance from the antennas to shower maximum, $X_{\rm max}$, is typically a few km, smaller than the characteristic length over which the shower develops, so that the Fraunhofer limit (relative to the shower dimensions) cannot be a good approximation. However, for zenith angles above $\theta=80^\circ$ the distances from antennas to shower maximum become more than 30 times larger than for vertical showers. 
The density of air at shower maximum for cosmic rays can be seen to span close to an order of magnitude as the zenith angle changes. Examples of the values of distance to $X_{\rm max}$, density at $X_{\rm max}$ and other relevant parameters for radio emission from cosmic ray showers for different zenith angles can be seen in Table\,\ref{tab:x_max_magnitudes}. The distance to shower maximum can become even larger than those in Table\,\ref{tab:x_max_magnitudes} if the antennas are located in the stratosphere as in the case of the ANITA balloon-borne detector\,\cite{ANITA:2010ect}, or in the case of detection from a satellite. It is clear that the Fraunhofer limit will be valid in some cases, but also that the conclusions obtained here would not apply for instance to near vertical cosmic ray showers detected with arrays of radio antennas. The density of air at $X_{\rm max}$ can be outside the range of Table\,\ref{tab:x_max_magnitudes} for instance for horizontal neutrino showers and for stratospheric showers. 
In Table\,\ref{tab:x_max_magnitudes} we also give $\Delta\rho$, defined as the change of density as a cosmic ray shower develops over six radiation lengths around shower maximum, three radiation lengths before and three after it. This region is responsible for the bulk of the contributions to the radio pulse. Changes in density in this region are of the order of $\pm 15 \%$ suggesting that reasonable results could be obtained assuming a constant density corresponding to the value at shower maximum. 

In this section we study the dependence of the frequency spectrum of the radio pulses  on the magnetic field intensity and on the density of air. 
With the newly developed ZHS program adapted to account for magnetic deflections, we have obtained the frequency spectrum of the radio emission in the Fraunhofer limit under different controlled conditions, to explore the properties of the pulses as the magnetic field and the air density are independently changed. We limit the discussion to the Frauhofer limit and to observers in the Cherenkov cone for simplicity. This exercise cannot be expected to give precise pulse predictions because the observers are not always placed at large distances to the shower and the decrease of the atmospheric density with height is not accounted for. However, as argued above, in many  experimental situations of interest the Fraunhofer condition is valid and the variation of air density is not too drastic so that the conclusions obtained here have very relevant implications for realistic showers.

\subsection{The effect of the magnetic field strength}

The new ZHS program allows us to obtain pulses from showers developing in air under the influence of magnetic fields of different intensities as can be naturally expected as showers develop in different Earth locations. To some extent it can be considered that the main effect of the magnetic field will be due to its component perpendicular to the shower direction. Depending on the relative orientation of the magnetic field and the shower direction (angle $\alpha$), the effective magnetic field strength will thus be reduced to $B \sin \alpha$. So it can also be expected the effective intensity to change from $0$ to its maximum value at any given location depending on $\alpha$~\cite{Zilles:2018kwq}. We will also use magnetic field intensities much higher than actual values at the Earth's surface that, although not very realistic, they will be shown to provide insight into the behaviour of showers developing in low density air. 

\begin{figure}[ht]
	\centering
	\subfigure{\includegraphics[width=0.45\linewidth]{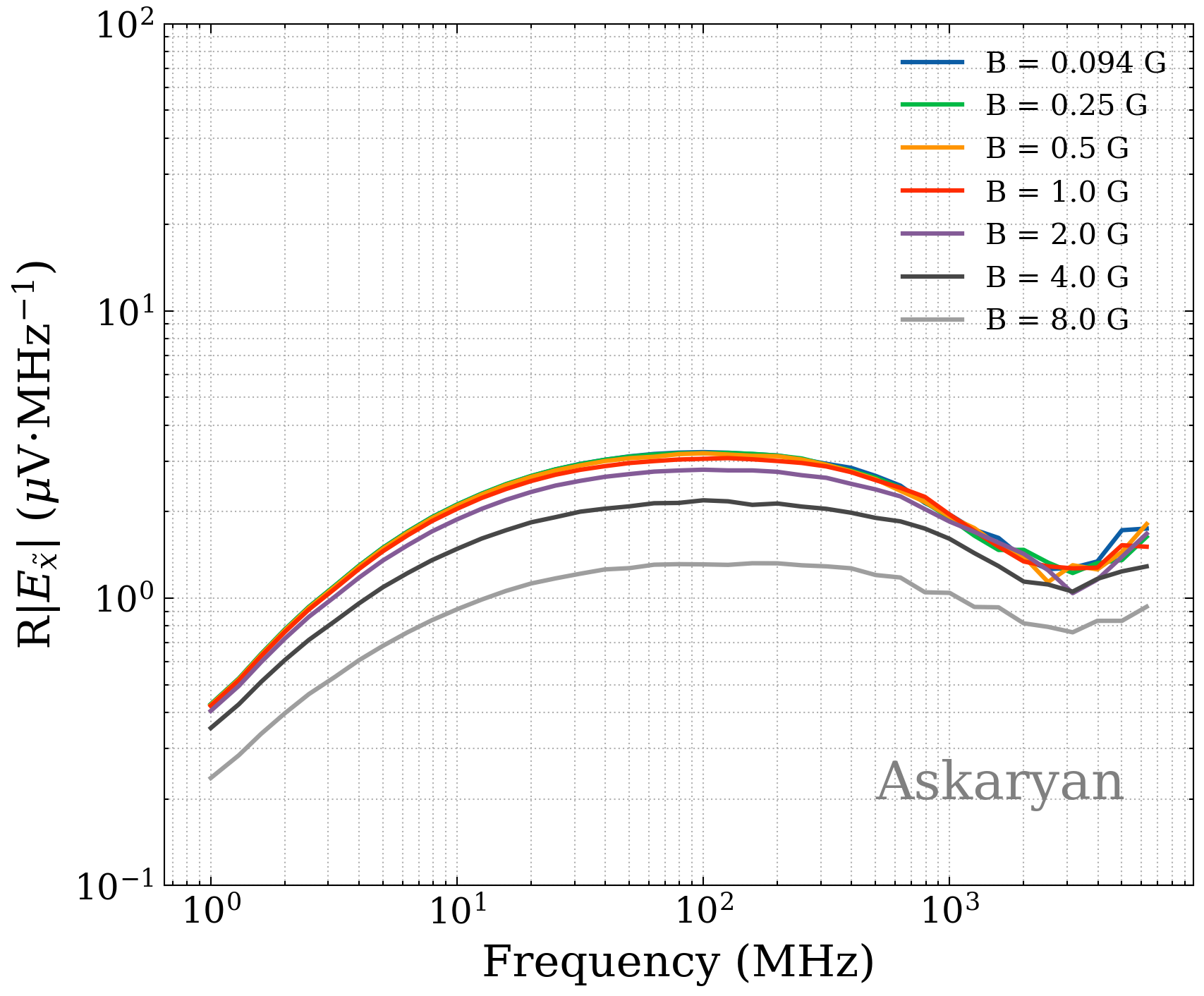}}
	\subfigure{\includegraphics[width=0.49\linewidth]{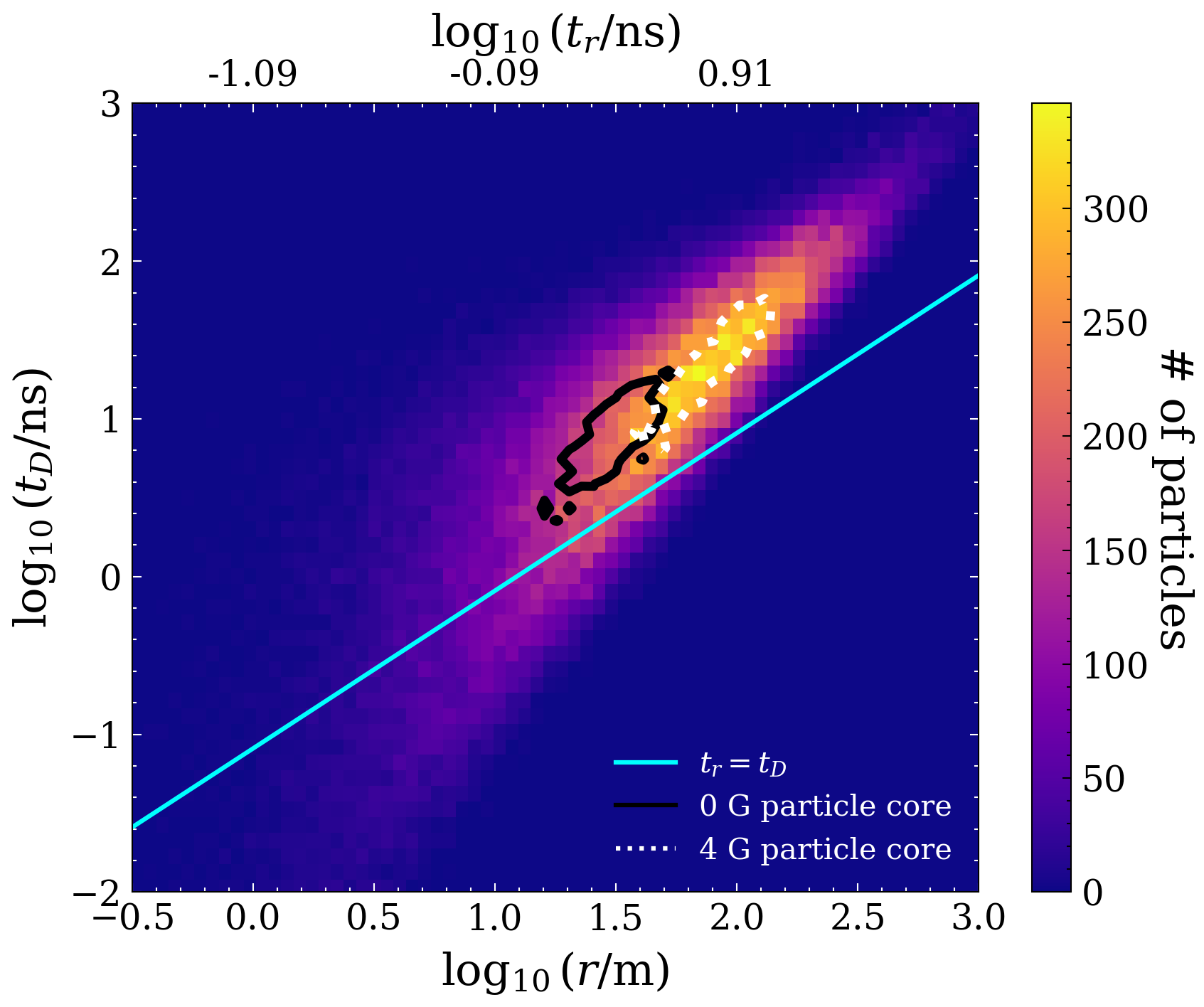}}
\caption{Left: Average amplitude of the frequency spectrum of the $E_{\tilde{x}}$ polarization (due to the excess charge) of the radio pulse for 50 electron showers of 100 TeV at sea level density with different values for the magnetic field as labeled. The geometry is that of Fig.\,\ref{fig:Geometry}. Right: Distribution of the shower particles in two-dimensional space given by the particle time delay and lateral position, as in Fig.\,\ref{fig:time_delay_hist} for an air shower near maximum, developing at sea level density and in the presence of a $B=4\,\mathrm{G}$ magnetic field perpendicular to the shower axis. The  contour lines are to guide the eye and indicate the peak of the distribution (containing about $10\%$ of the total) for this shower (white) compared to the equivalent histogram in the case of no magnetic field (very similar to that in Fig.~\ref{fig:time_delay_hist} for $B=0.5$~G). The light blue line gives the region below which the maximum value of $t_r$ exceeds $t_D$.}
\label{fig:magnetic_field_strength_Askaryan}
\end{figure}
We have simulated seven sets of 50 electron showers of 100 TeV each in air of sea-level density with seven different magnetic field intensities and the geometry sketched in Fig.\,\ref{fig:Geometry}. We scale a reference value of $B_0=0.5\,\mathrm{G}$ with a factor $f_B=B/B_0=0.2,0.5,1,2,4,8,16$, to give $B=0.1, 0.25, 0.5, 1, 2, 4, 8$~G. The results for the spectral amplitude of the $E_{\tilde{x}}$ component, due to the Askaryan effect, are displayed in the left panel of Fig.~\ref{fig:magnetic_field_strength_Askaryan}. The plot clearly shows that the emitted pulse is 
independent of the magnetic field for $B<1\,\mathrm{G}$. For larger values the spectral features remain very similar yet the amplitude drops gradually. This is not surprising, the excess charge is not expected to be directly affected by the magnetic field intensity. The deviations that are shown for values of $B>1$~G can be interpreted as a loss of coherence induced by the magnetic field. The curvature affects both the particle delays $t_D$ because as the particle rotates it lags relative to the plane moving along the shower axis at the speed of light, and also the geometric delays, $t_r$, due to the lateral position of the particles because the magnetic field moves particles away from the shower axis. The extra delays due to the magnetic field only become relevant compared to the typical delays of the showering process for a magnetic field exceeding $\sim 1\,\mathrm{G}$ for air at sea level density. 

The right panel of Fig.~\ref{fig:magnetic_field_strength_Askaryan} displays a two-dimensional histogram of the time delays of particles at the maximum of a shower developing in air at sea level density and in the presence of a perpendicular magnetic field of strength $B=4\,\mathrm{G}$. By comparison to the histogram shown in Fig.\,\ref{fig:time_delay_hist} for the case $B=0.5\,\mathrm{G}$ it can be seen that the bulk of the particle distribution has shifted to the upper right corner (contour lines containing the bulk of the particles in both cases are superimposed to visualize the effect). Both $t_D$ and $t_r$ can be seen to have increased because of the intense magnetic field applied. 
When magnetic delays become relevant, particles that acquire large time delays due to the magnetic field cease to contribute in a fully coherent fashion to the emission, thus reducing the resulting electric field amplitude. For very high frequencies in the GHz region, only a small number of particles with delays below a fraction of a nanosecond contribute coherently. They populate the lower left corner of the particle distribution displayed in the left panel of  Fig.~\ref{fig:magnetic_field_strength_Askaryan}, and they typically have high energy (high rigidity). The trajectories of these particles are less curved in the magnetic field and their time delays are less affected, explaining why the frequency spectrum, in the GHz frequency region, drops less in this region. 
Regarding the shape of the spectrum it can be seen that even at $B=4\,\mathrm{G}$, the spectral shape of the $E_{\tilde{x}}$ component in the left panel of Fig.~\ref{fig:magnetic_field_strength_Askaryan} is very much the same as for lower values of $B$. However, for $B=8\,\mathrm{G}$ it can be seen that the spectrum ceases to increase linearly with frequency at lower frequencies than in the case $B=0.5\,\mathrm{G}$. This is again attributed to the fact that some particles (in the top right of the histogram) acquire very large delays and stop contributing fully coherent also at lower frequencies. 

\begin{figure}[ht]
	\subfigure{\includegraphics[width=0.49\linewidth]{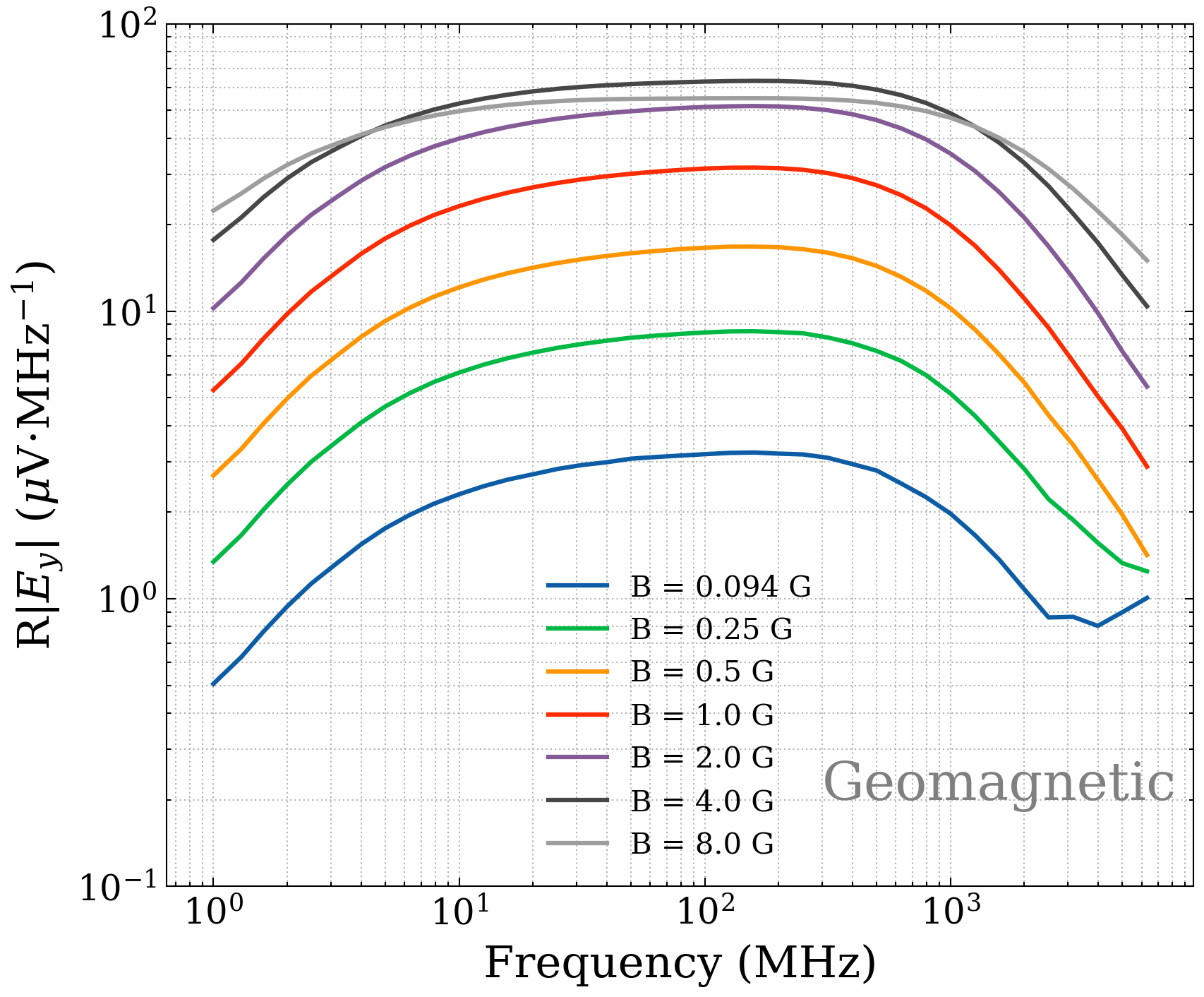}}
	\subfigure{\includegraphics[width=0.49\linewidth]{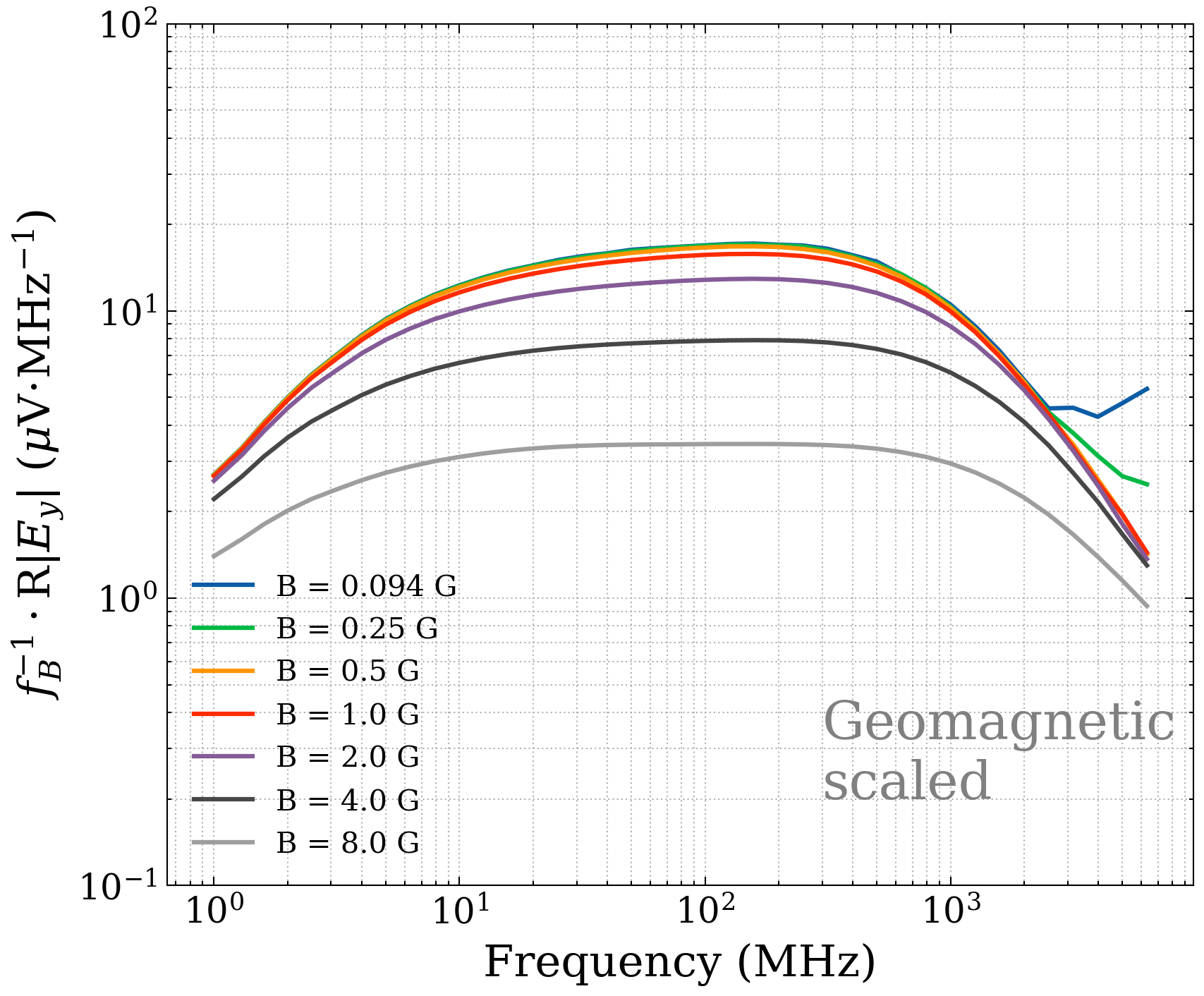}}
\caption{Left panel: Average amplitude of the frequency spectrum of the $E_y$ polarization (due to the geomagnetic effect) of the radio pulse for 50 electron showers of 100 TeV at sea level density for each value of the magnetic field intensity $B$ as labeled. The geometry is that of Fig.\,\ref{fig:Geometry}.
In the right panel, the amplitude of $E_y$ has been divided by the scaling factor $f_B=B/B_0$ with $B_0=0.5\,\mathrm{G}$ to clearly display the scaling behavior and the deviations from a linear scaling with $B$ that appear when $B>1\,\mathrm{G}$.}  
\label{fig:magnetic_field_strength_polarizations}
\end{figure}
The $E_y$ component (due to the geomagnetic effect) can be seen in the left panel of  Fig.~\ref{fig:magnetic_field_strength_polarizations} to increase as the magnetic field intensity $B$ grows. The increase appears to be  linear in $B$ up to about $B\sim 1\,\mathrm{G}$. For $B>1\,\mathrm{G}$ the amplitude increase is mitigated, and for $B> 4\,\mathrm{G}$ the amplitude does not increase significantly, appearing to reach some kind of "saturation" or even to drop slightly. This can be better appreciated in the right-hand panel where the normalization of each curve is divided by the scaling factor $f_B$. The curves for $B=0.094,0.25,0.5$ and $1\,\mathrm{G}$ can hardly be distinguished. For $B=1,\,2,\,4\,\mathrm{G}$ the amplitudes of the obtained pulses scaled down by $f_B$ are respectively reduced by $6.3,23.1,52.8\%$ relative to the reference amplitude for $B=0.5$~G.  The deviations from the perfect scaling at high magnetic field intensities can be understood again with the argument that the increased time delays and transverse deviations due to the Lorentz force reduce the number of particles contributing coherently. 
The scaling of the electric field amplitude with magnetic intensity has been postulated for long to be proportional to $B \sin \alpha $~\cite{PierreAuger:2016vya, falcke2005detection, aab2014probing, ardouin2009geomagnetic} because the Lorentz force is proportional to ${\boldsymbol v} \cross {\bf B}$. This has been tested both with simulations and in experiments~\cite{PhysRevLett.116.141103}. For values of $B<0.5\,\mathrm{G}$ and air at sea level density this scaling is accurate to better than $\sim 2\%$ in the Fraunhofer limit. The scaling holds for $B=1\,\mathrm{G}$ to $\sim 8\%$ accuracy and worsens for $B>1\,\mathrm{G}$.

\subsection{The effect of air density on radio emission}

We now describe the effect of changing the density of air which is more complex. For a start it is coupled to a change in the Cherenkov angle through the dependence of the refractive index on density. 
For a low density medium such as air, the refractive index is very close to 1 and in turn the Cherenkov angle is small and can be approximated as: 
\begin{equation}
    \cos \theta_C = \frac{1}{n} \simeq 1 - \frac{\theta_C^2}{2} 
    \quad \Rightarrow \quad \theta_C \simeq \sqrt{2\left(\frac{n-1}{n}\right)}.
\label{eq:cher_angle}
\end{equation}
As the refractivity $(\mathcal{R}=n-1)$ scales approximately linearly with density $\rho$, the refractive index can be written as, 
\begin{equation}
    n-1 = a \rho = \mathcal{R}_0\frac{\rho}{\rho_0},
\end{equation}
where $\mathcal{R}_0$ and $\rho_0$ are the refractivity and density at sea level. Using Eq.\,(\ref{eq:cher_angle}), the Cherenkov angle is expected to approximately scale with the square root of the density:
\begin{equation}
    \theta_C = \sqrt{2\,\frac{\mathcal{R}_0}{n}\frac{\rho}{\rho_0}}\simeq\sqrt{2\,\mathcal{R}_0\frac{\rho}{\rho_0}}.
\label{eq:theta_var_n}
\end{equation}
%
\subsubsection{The effect of changing the refractive index alone}

In view of Eq.\,(\ref{eq:theta_var_n}) it is interesting to first study the effect of changing the refractive index alone, in spite of this being unphysical in the sense that a change in the refractive index is coupled to a change of air density. For this purpose, with ZHS we have simulated five sets of 50 electron showers each, propagating in air with five different refractive indices corresponding to Cherenkov angles $\theta_C=1.4^\circ$, $0.99^\circ$, $0.7^\circ$, $0.5^\circ$ and $0.35^\circ$\,\footnote{The refractive indices correspond to the densities of air chosen later in this section, that is, the refractive index at sea level density of $\rho_0=1.2\times 10^{-3}\,{\rm g\,cm^{-3}}$ and those corresponding to $\rho_0$ reduced by factors $f_\rho=2, 4, 8$ and $16$.}, keeping the air density constant at the sea level value of $\rho_0=1.2 \times 10^{-3}\,{\rm g\,cm^{-3}}$. The magnetic field intensity is fixed at $B=0.5\,\mathrm{G}$, with the  geometry shown in Fig.~\ref{fig:Geometry}.

\begin{figure}[ht]
\centering
\subfigure{\includegraphics[width=0.49\linewidth]{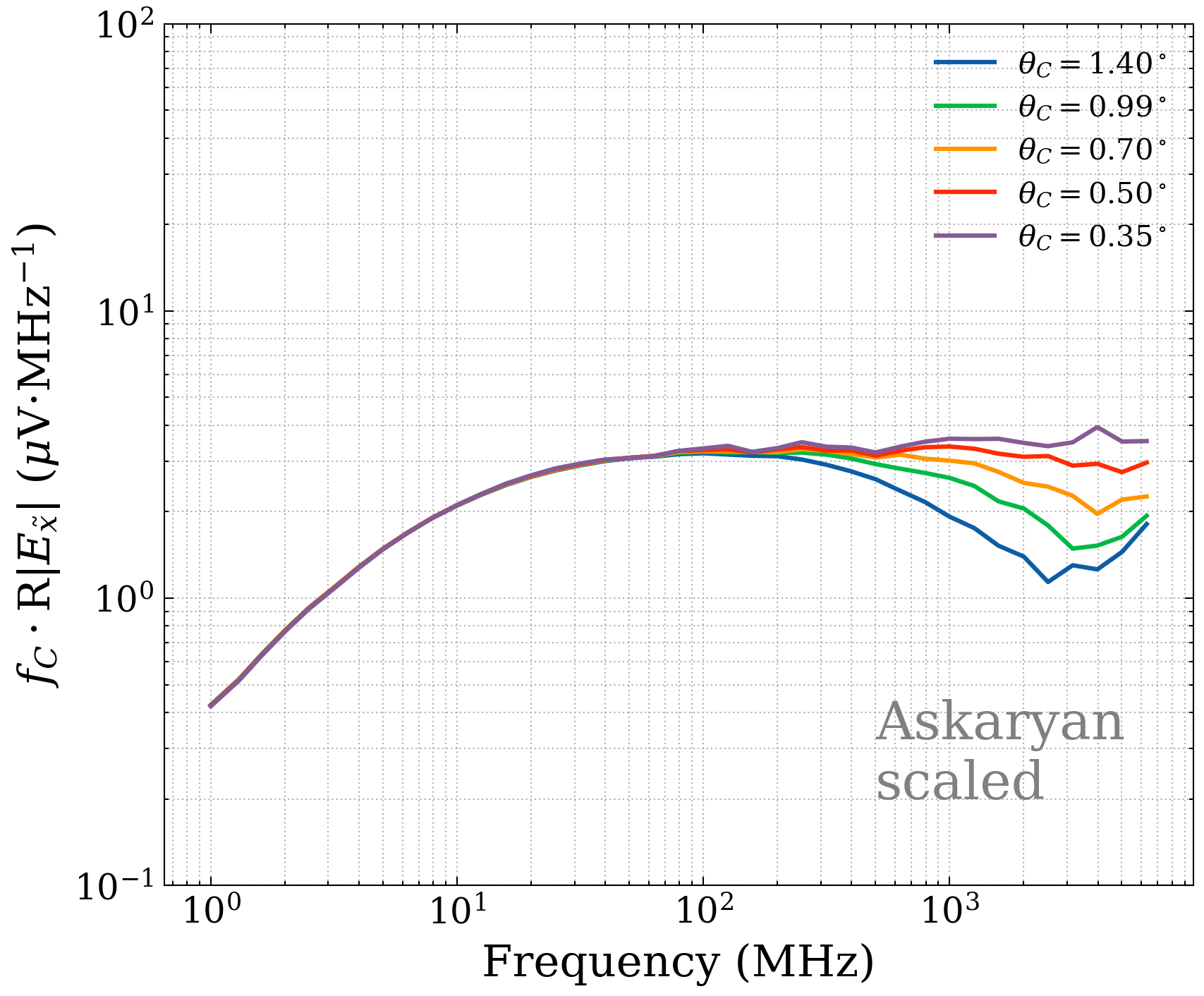}}
\subfigure{\includegraphics[width=0.49\linewidth]{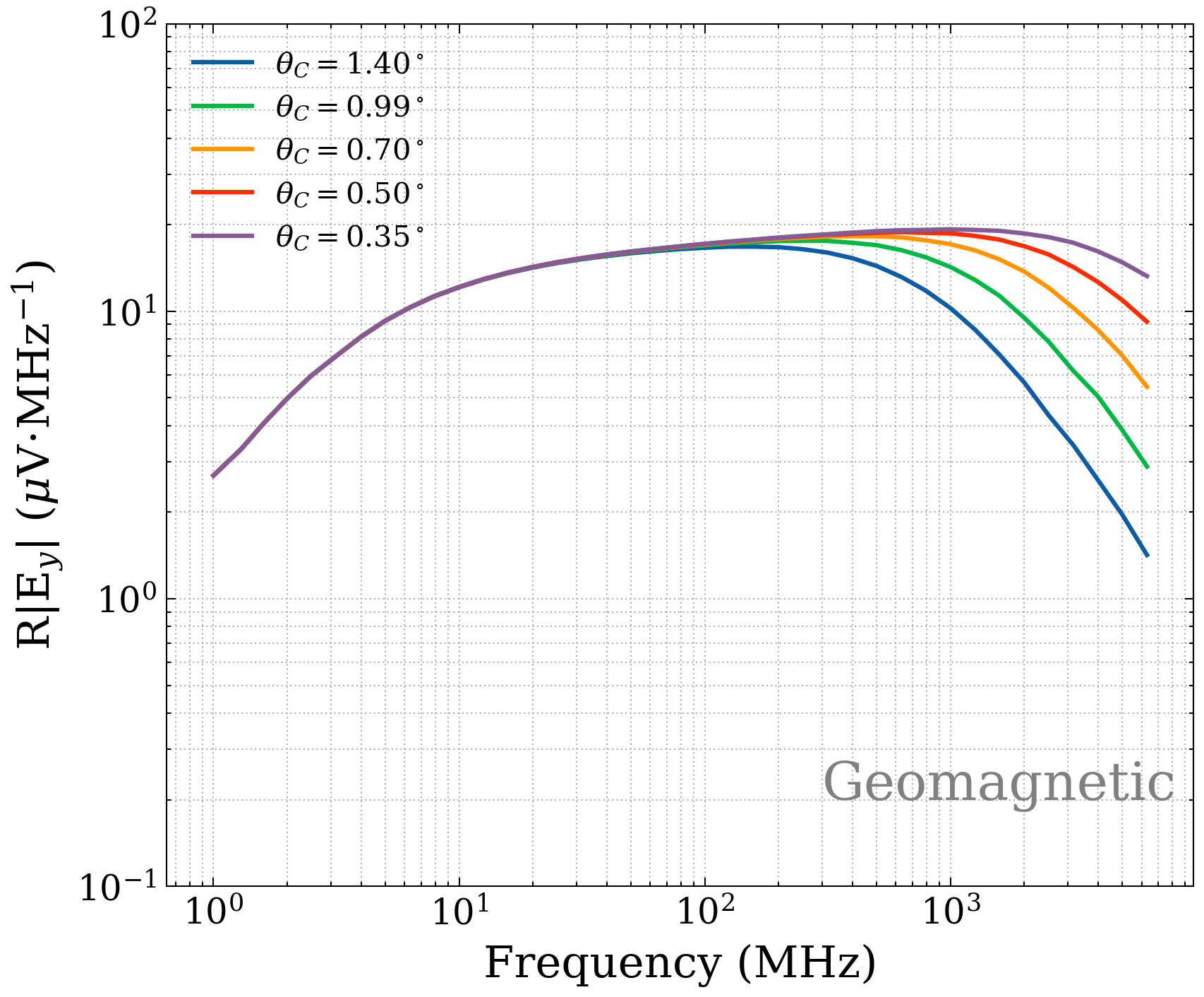}}
\caption{Average frequency spectrum of the radio pulses of 50 showers initiated by a 100 TeV electron developing in air at constant sea level atmospheric density with different refractive indexes. The magnetic field intensity is fixed to $B=0.5\,\mathrm{G}$. The observer is placed at the Cherenkov angle in the direction of the magnetic field as in Fig.\,\ref{fig:Geometry}. In the left panel the $E_{\tilde{x}}$ component of the field is shown, determined mainly by the Askar'yan emission mechanism.  $E_{\tilde{x}}$ has been scaled with a factor $f_C=\sin\theta_{C_{_0}}/\sin\theta_C$ where $\sin\theta_{C_{_0}}$ is the sine of the Cherenkov angle corresponding to sea level refractive index $\theta_{C_{_0}}\simeq 1.4^\circ$. In the right panel we show the $E_y$ component dominated by the geomagnetic mechanism.}
\label{fig:cte_rho_var_n_scaling}
\end{figure}

The results for the average Fourier amplitudes of the $E_{\tilde{x}}$ and $E_y$ pulse components are respectively shown in the left and right panels of Fig.~\ref{fig:cte_rho_var_n_scaling}. 
In this geometry, the $E_{\tilde{x}}$ component is due to the charge excess and the $E_y$ component, parallel to the transverse current, is attributed to the geomagnetic effect. 
%
%
Each curve in the left panel of  Fig.~\ref{fig:cte_rho_var_n_scaling} describing the charge excess component shows the amplitude divided by the factor $f_C=\sin\theta_{C_0}/\sin\theta_C$ with $\theta_{C_0}=1.4^\circ$. With this scaling factor, all the spectra align quite precisely for frequencies below $\sim 100$~MHz. This is easy to interpret. The electric field intensity due to the charge excess is proportional to ${\boldsymbol v}_{\perp} $, the particle velocity projected onto the perpendicular to the observer direction in the $\boldsymbol{\hat{v}}\boldsymbol{\hat{k}}$ plane as can be seen in Eq.\,(\ref{eq:ZHS}), where $v_\perp = v \sin \theta$ and $\theta$ is the observation angle. The electric field due to the Askar'yan emission is thus expected to be proportional to $\sin \theta_C$ for observation from the Cherenkov direction whatever the azimuthal observation angle $\phi$. This has been checked explicitly in the simulations. 
 
The $E_{y}$ component is mainly due to the geomagnetic emission mechanism and is shown without any scaling factor in the right panel. Following the same reasoning above, the transverse current induced by the magnetic field is perpendicular to ${\boldsymbol v}$, in the direction ${\boldsymbol v} \cross {\boldsymbol B}$. For an observer in the $\boldsymbol{\hat{x}}\boldsymbol{\hat{z}}$ plane in Fig.\,\ref{fig:Geometry}, the transverse current is already perpendicular to the observing direction and no projection factor is needed. If the observer was instead in the  $\boldsymbol{\hat{y}}\boldsymbol{\hat{z}}$ plane ($\phi=90^\circ$) a projection factor of $\cos\theta$ would be needed for the transverse current\footnote{For an arbitrary observation azimuth, $\phi$, the projection angle would be smaller than $\cos\theta$.}. In any case, as long as the observer is close to the Cherenkov angle,  $\theta \sim \theta_C = 1.4^\circ$ the projection factor is nearly one. The geomagnetic component for observation near the Cherenkov direction is to a very good approximation independent of the refractive index as shown in the right panel of Fig.\,\ref{fig:cte_rho_var_n_scaling}. It can also be appreciated that the high-frequency components become less suppressed as the refractive index, $n$, is lowered. This is due to the lateral spread, inducing a delay $t_r\simeq n r \sin \theta \cos \phi$ (Eq.~\ref{eq:zhs_formula_cher}) which can dominate over the delays, $t_D$, at frequencies exceeding 200 MHz as discussed in Section~\ref{AskaryanInAir}. If the refractive index (and hence the Cherenkov angle) is reduced, the $t_r$ delays, scaling linearly with $n \sin \theta$, are also reduced and the emission will be more coherent at a fixed frequency as shown in the right panel of Fig.\,\ref{fig:cte_rho_var_n_scaling}. 


\subsubsection{The effect of changing the density alone}

\begin{figure}[ht]
\centering
\subfigure{\includegraphics[width=0.49\linewidth]{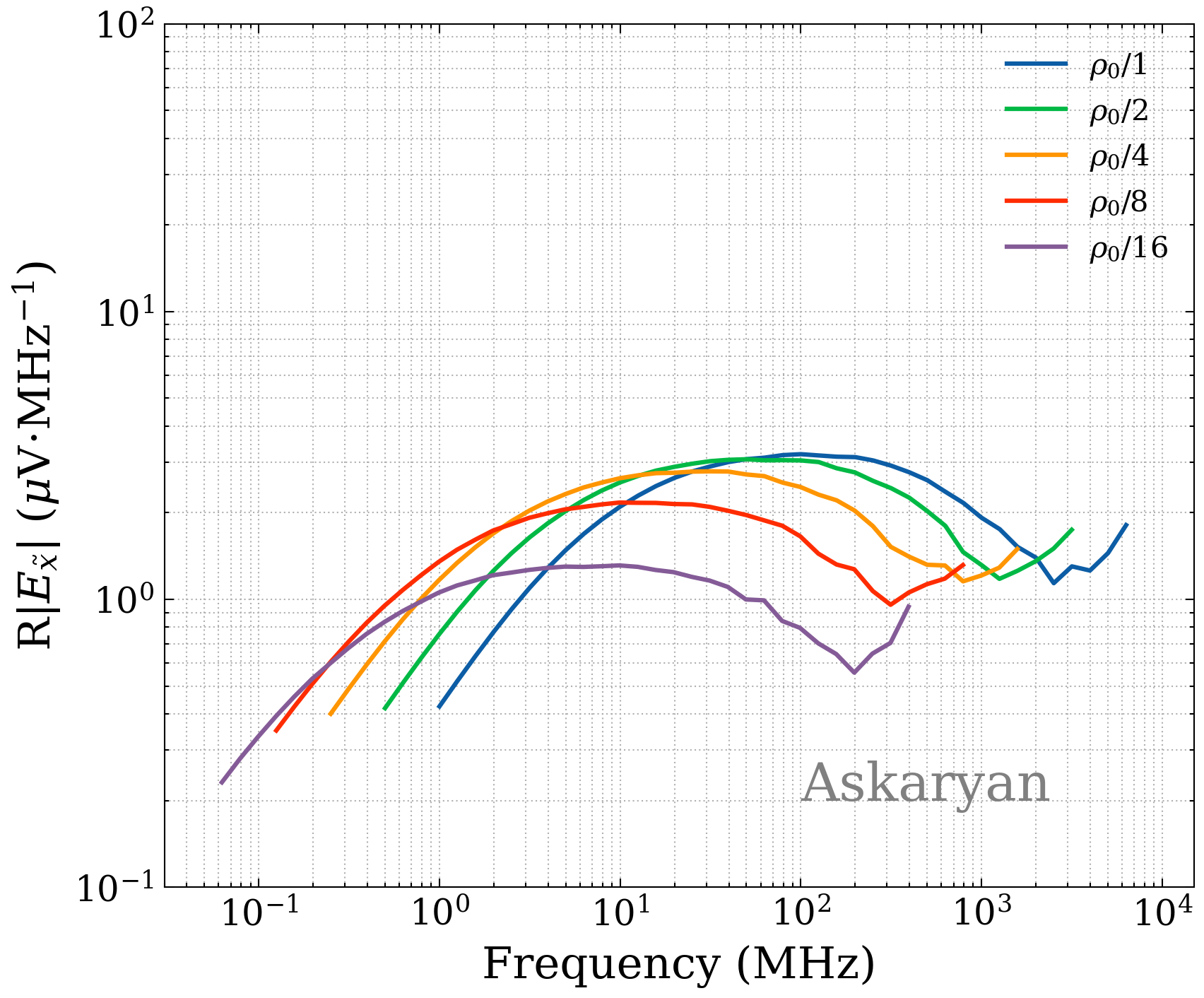}}
\subfigure{\includegraphics[width=0.49\linewidth]{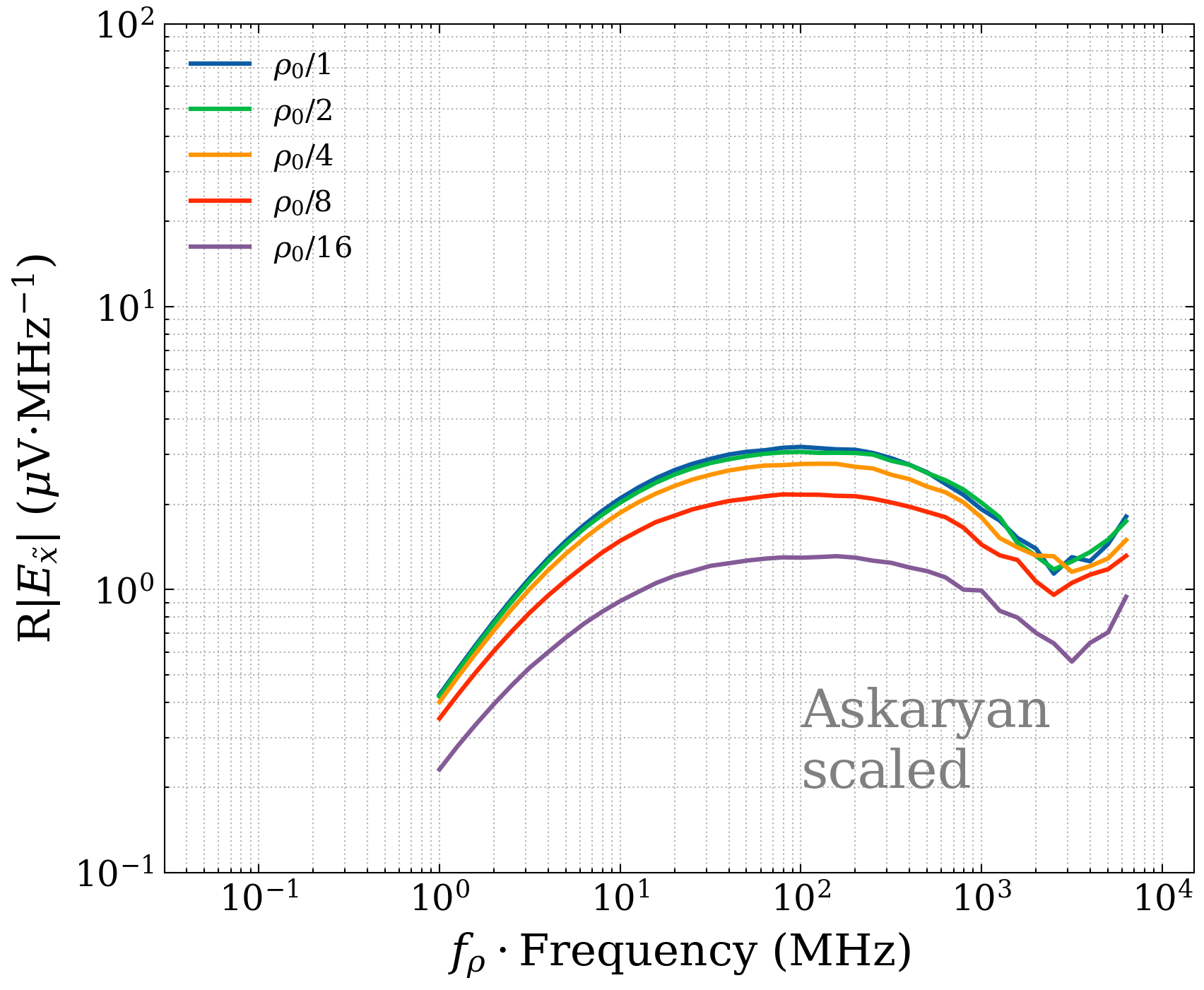}}
\subfigure{\includegraphics[width=0.49\linewidth]{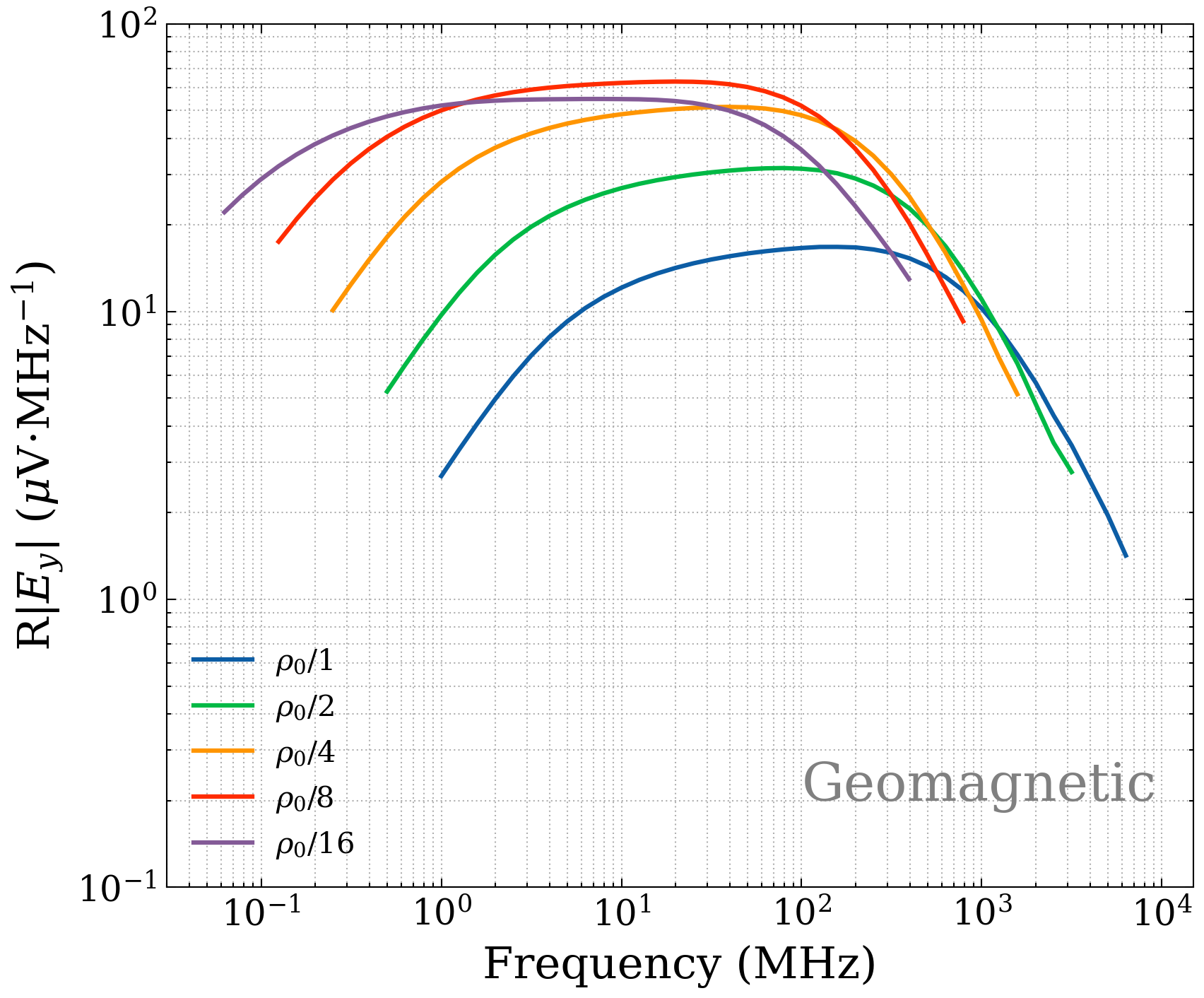}}
\subfigure{\includegraphics[width=0.49\linewidth]{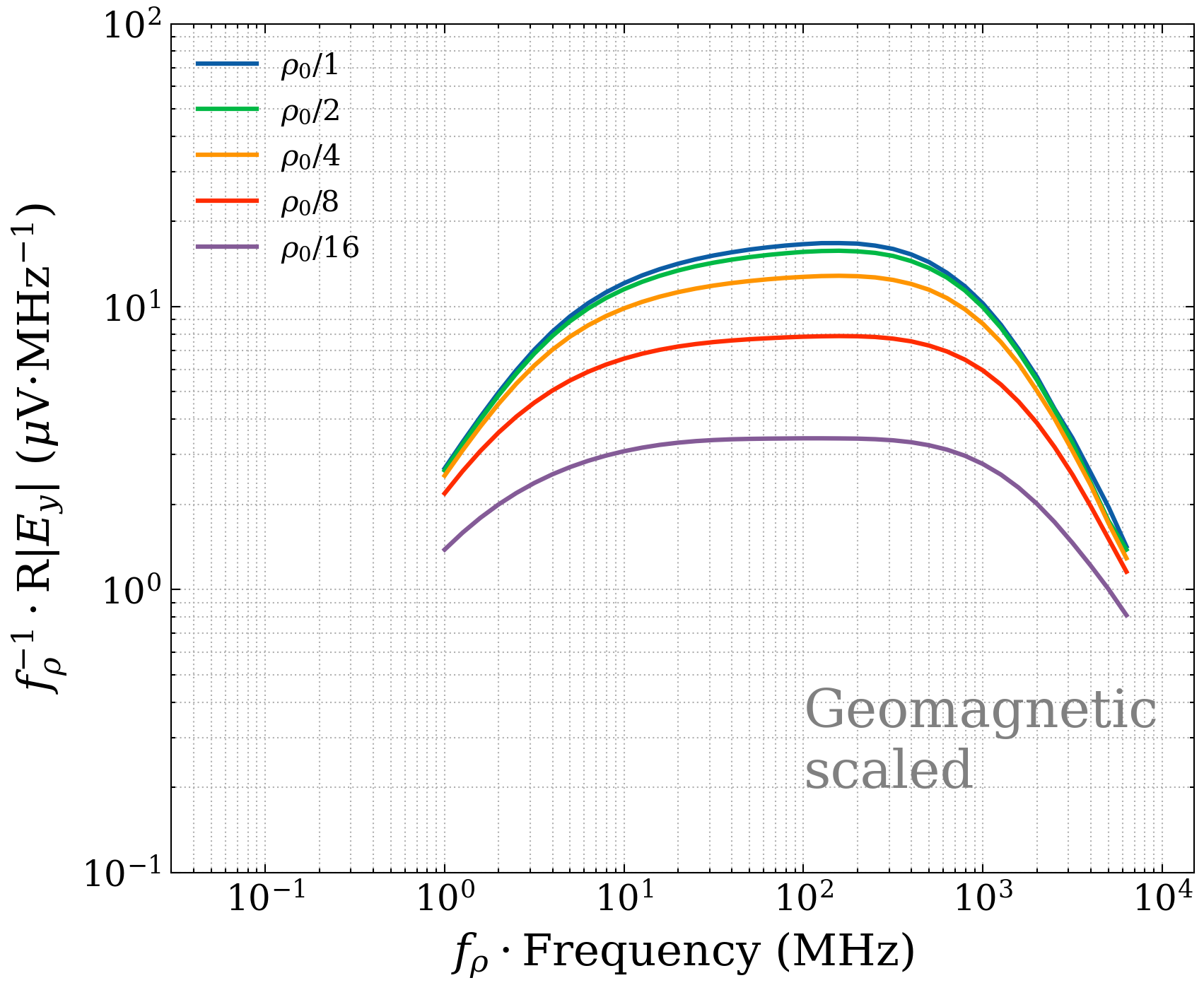}}
\caption{Average frequency spectrum of the radio pulses of 50 showers initiated by a 100 TeV electron developing in air at five different (constant) densities as labelled. The refractive index is artificially kept at the sea level value for all curves. The magnetic field intensity is $B=0.5\,\mathrm{G}$ in the $\boldsymbol{\hat{x}}$ direction and the observer is placed at the Cherenkov angle as shown in Fig.\,\ref{fig:Geometry}. In the top panels the $E_{\tilde{x}}$ component of the field is shown, determined mainly by the Askaryan emission mechanism. In the bottom panels the $E_y$ component, dominated by the geomagnetic mechanism, is displayed. The plots on the left show the obtained frequency spectra, while on the right the same curves are shown replacing the frequency axis in the plot with the frequency multiplied by the density scaling factor $f_\rho=\rho_0/\rho$ with $\rho$ the density of air at sea level. Besides, in the bottom right plot the $E_y$ component of the electric field is multiplied by the same factor $f_\rho$ as explained in the text.}
\label{fig:var_rho_cte_n_ask} 
\end{figure}

We now leave the refractive index and Cherenkov angle fixed and perform simulations changing the air density to $1.2$, $0.6$, $0.3$, $0.15$ and $0.075$~g~cm$^{-3}$, corresponding to the sea level air density $\rho_0$ respectively reduced by factors $f_\rho=1, 2, 4, 8, 16$. We will refer to $f_\rho=\rho_0/\rho$ as the "density reduction factor". Even ignoring magnetic deflections, reducing the density by a factor $f_\rho$ has a direct impact on electromagnetic shower development and radio emission, increasing both the physical dimensions of the shower and the time delays by the same factor. In the Cherenkov direction the total time delay of a signal from a given track is split into the particle delay relative to a plane travelling at the speed of light, $t_D$, and the signal delay due to its lateral position, $t_r$. Both scale with the distance travelled by particles so that the total delay $t_r+t_D$ will also increase linearly with the inverse of the density. An increase of the total time delay can be compensated for by an equivalent decrease in frequency to give a similar phase factor in Eq.\,(\ref{eq:zhs_formula_cher}). As a result, if a given spectral feature occurs at a particular frequency $\nu$ in a shower developing in air at sea level density $\rho_0$, it will also show up at a frequency $\nu/f_\rho$ scaled down with a factor $f_\rho$ for a density $\rho_0/f_\rho$. This can be made apparent if the spectral amplitudes for both showers are plotted versus the frequency scaled up with $f_\rho$. This is clearly shown in the upper plots of Fig.~\ref{fig:var_rho_cte_n_ask} comparing the spectral amplitude of $E_{\tilde{x}}$ as obtained in the simulations for the five air density values (top left panel) to the  plots in the top right panel where the frequency is scaled with $f_\rho$. As can be seen, the spectral features line up, but the normalization does not quite match for all curves as discussed below. 

The normalization of the electric field amplitude is proportional both to the frequency $\nu$ and to the projected tracklength of the particle trajectories $v_\perp\delta t$, as can be read from Eq.\,(\ref{eq:zhs_formula_cher}). 
As the density drops by a factor $f_\rho$, the tracklength increases by $f_\rho$, so that when we consider the low frequency part of the spectrum, proportional to $\nu$ (all particles emitting coherently), the amplitude can be seen to increase linearly with $f_\rho$ for the curves with $f_\rho=1$ and 2 (top left panel of Fig.~\ref{fig:var_rho_cte_n_ask}). In the top right panel an abscissa of value $\nu_0$ corresponds to a frequency $\nu_0$ for $f_\rho=1$ but to $\nu_0/2$ for $f_\rho=2$. This artificial frequency shift compensates for the increase in tracklength due to the lower density. This simple scaling is broken progressively as $f_\rho$ becomes 4, 8 and 16, marking the transition where the relative delays and position shifts due to deflections in the magnetic field gradually become more relevant. As the density is lowered, the interaction length increases, particles have more time between successive interactions while keeping the same radius of curvature and hence deflect more. As a consequence coherence loss due to magnetic effects are relatively more important.

The spectral amplitude of the geomagnetic component of the field, $E_y$, is displayed in the bottom left panel of Fig.\,\ref{fig:var_rho_cte_n_ask} which shows an increase in amplitude as the density is reduced. Only in the very low frequency region, below 1 MHz/$f_\rho$, where the field is accurately proportional to $\nu$, the increase in amplitude of the geomagnetic component scales with $f_\rho^2$. As the density drops, besides the increase of tracklength, there is another effect that has an impact on the geomagnetic component. The transverse current that develops also increases due to the Lorentz force acting over longer time because the typical distance between particle interactions has increased. The transverse current is proportional to an effective drift velocity in the transverse plane~\footnote{Similar to Drude's model.} that grows with $f_\rho$~\cite{Scholten:2007ky}. This suggests a scaling of the amplitude with $f_\rho$ that cannot be compensated with the scaling of frequency as in the case of tracklength. In the bottom right panel of Fig.~\ref{fig:var_rho_cte_n_ask} we have plotted the spectra scaling the frequency with $f_\rho$, in the same way as for the $E_{\tilde{x}}$ component, so that the spectral features line up. In addition the amplitude of the cases with reduced density has been reduced by $f_\rho$ to account for the increased transverse current. In the bottom right panel of Fig.\,\ref{fig:var_rho_cte_n_ask} the electric field amplitude for sea-level density reduced by a factor of two lies below the curve for sea level density ($\sim 6\%$ below in the central region). For $f_\rho=4$ ($\rho=0.3$~g~cm$^{-3}$) the curve shifts down by about $\sim 23\%$ and for larger values of $f_\rho$ the scaling is lost and the amplitudes display larger drops. The reason is that as the density decreases particles acquire higher transverse velocities which inevitably increases their time delays so that the fraction of particles contributing coherently reduces relative to the simple scaling law.

Finally, it is also interesting to see that the pattern in the frequency spectra displayed by the family of curves for the $E_y$ component is practically identical to the pattern displayed by $E_{\tilde{x}}$. 

\begin{figure}[ht]
\centering
\subfigure{\includegraphics[width=0.49\linewidth]{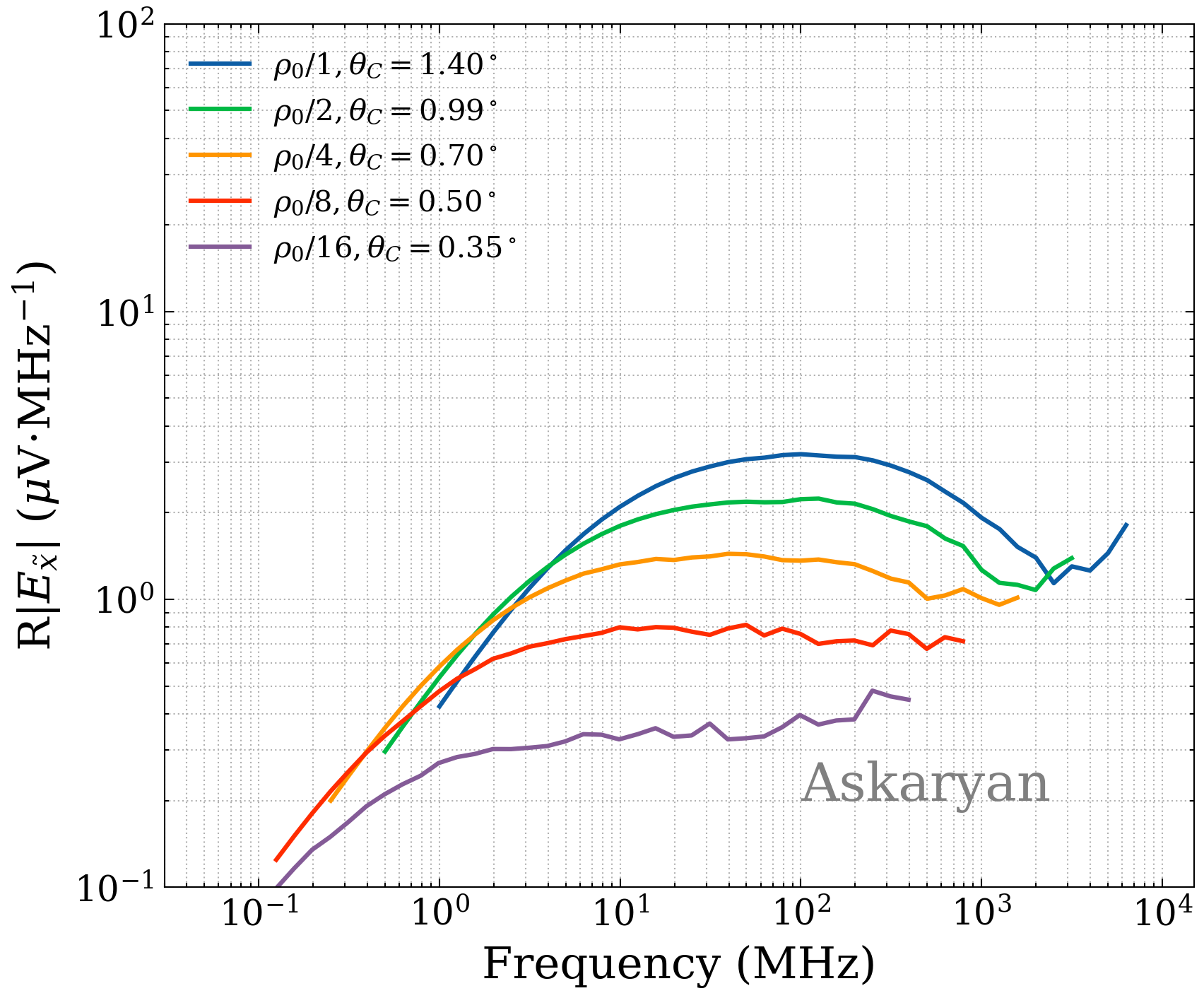}}
\subfigure{\includegraphics[width=0.49\linewidth]{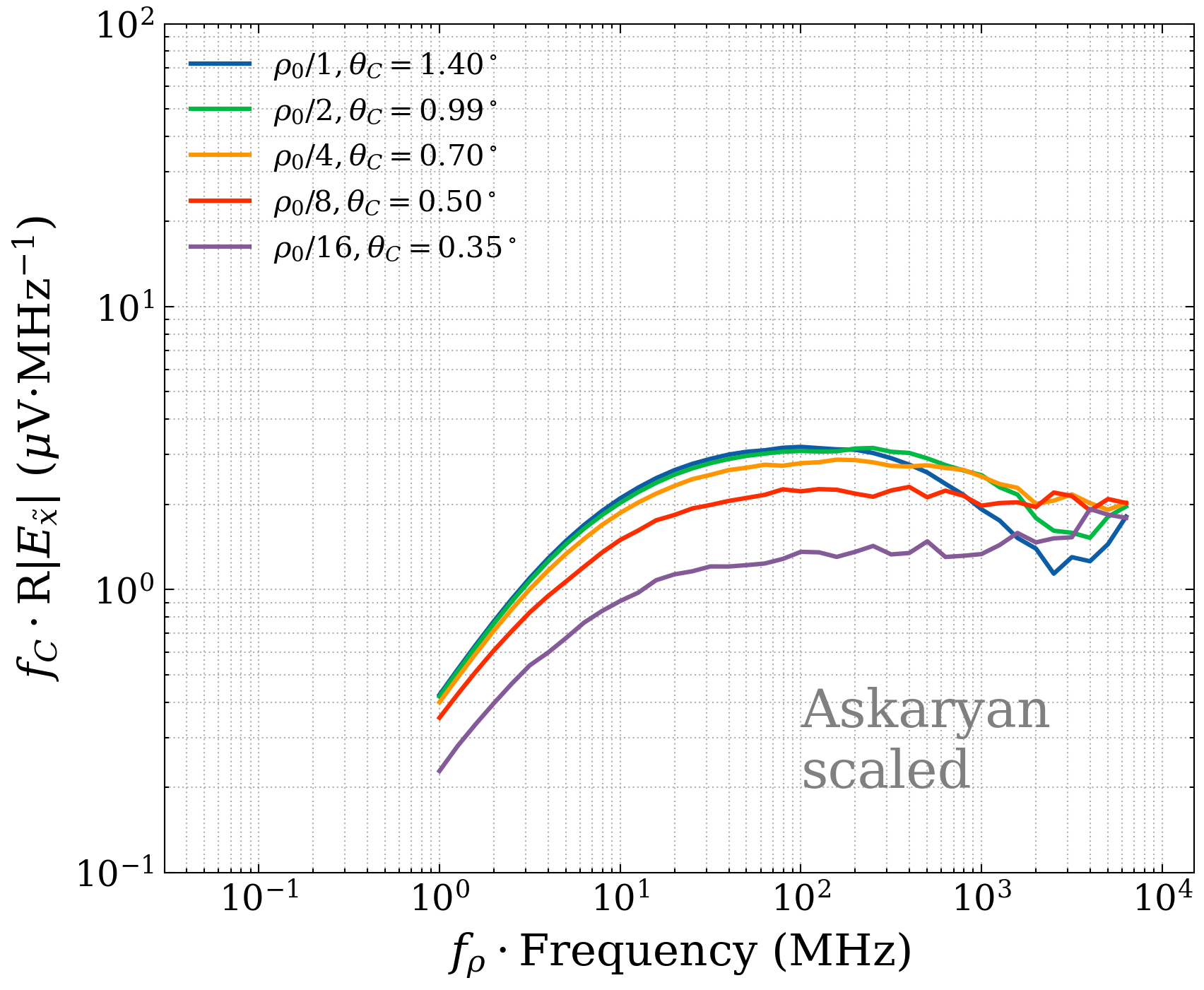}}
\subfigure{\includegraphics[width=0.49\linewidth]{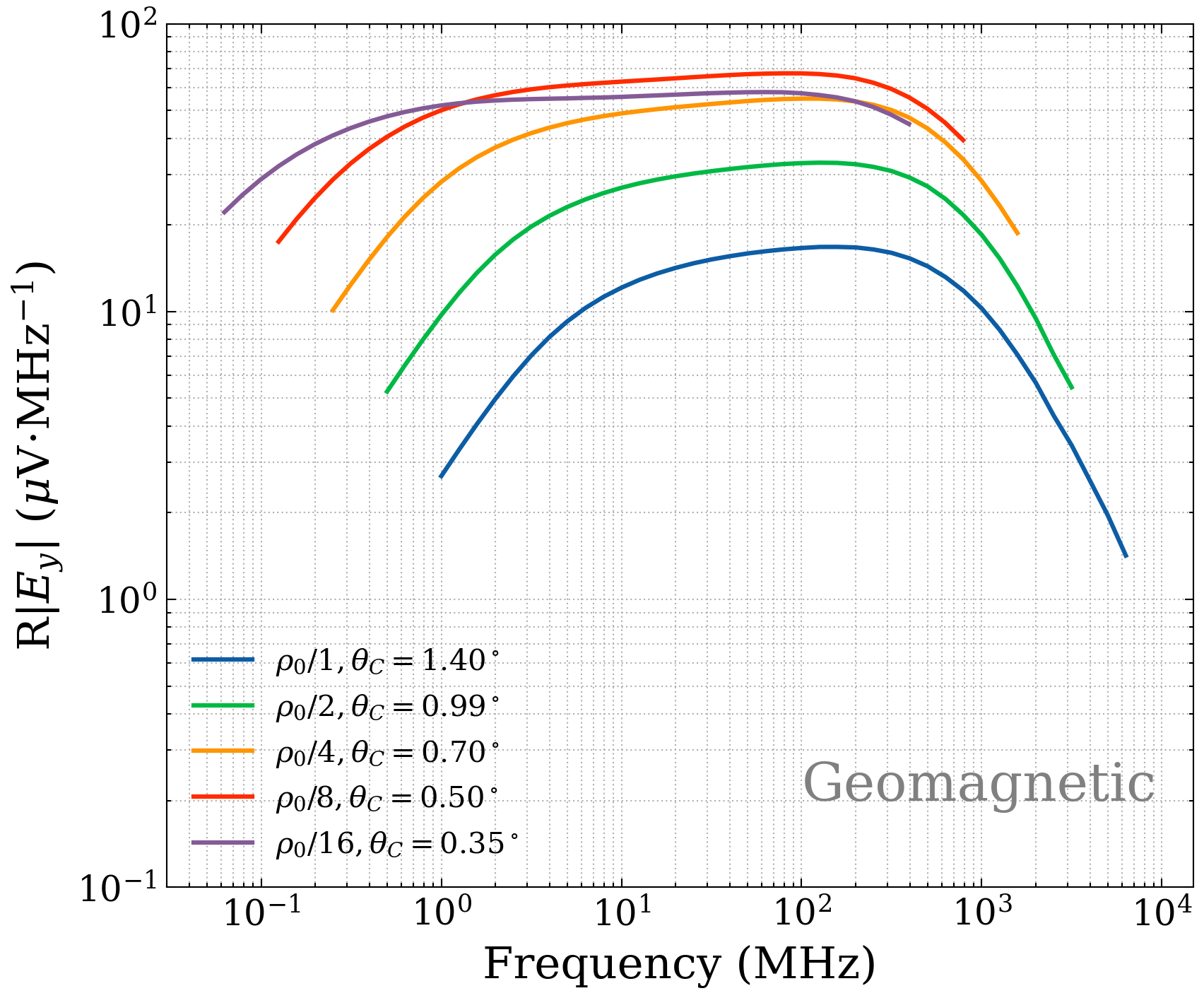}}
\subfigure{\includegraphics[width=0.49\linewidth]{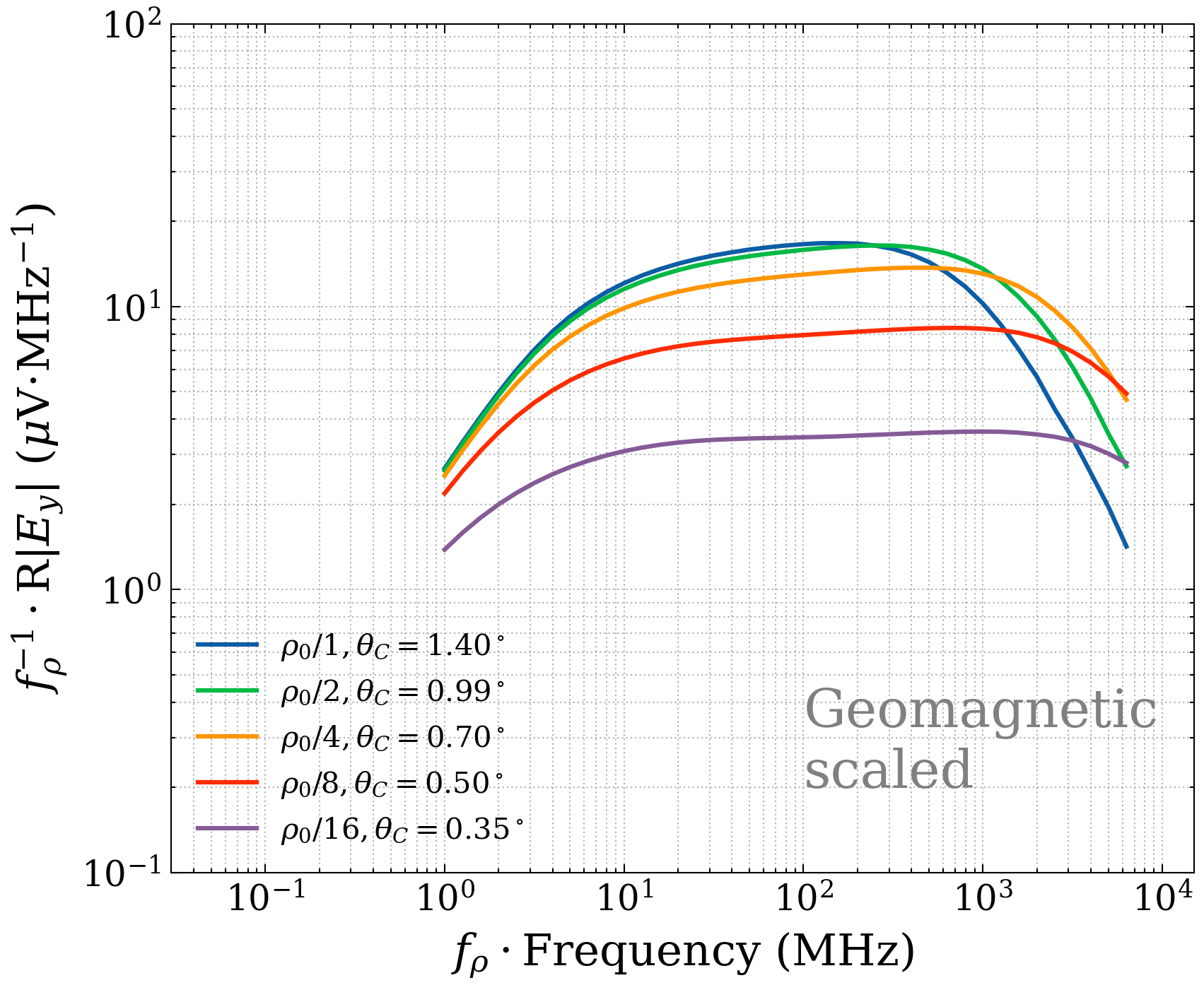}}
\caption{Average frequency spectrum of the radio pulses of 50 showers initiated by a 100 TeV electron developing in air at different air densities with corresponding different refractive indexes. The magnetic field intensity is $B=0.5\,\mathrm{G}$. The observer was placed at the Cherenkov angle in the direction of the magnetic field (see Fig.\,\ref{fig:Geometry}). In the left panel the $E_{\tilde{x}}$ component of the field is shown, determined mainly by the Askaryan emission mechanism. In the right panel we show the $E_y$ component dominated by the geomagnetic mechanism. Frequencies and the amplitude of the $E_y$ component of the electric field have been scaled accordingly with the density factor as explained in the text. $E_{\tilde{x}}$ is scaled with $\sin\theta_{C_{_0}}/\sin\theta_C$ where $\sin\theta_{C_{_0}}$ is the sine of the Cherenkov angle corresponding to sea level refractive index.}
\label{fig:var_rho_var_n_cher_scaling}
\end{figure}

\subsubsection{The effect of consistent changes in the air density and the refractive index}

We are now ready to deal with the physical case of both the Cherenkov angle and the density changing consistently. The results are shown in Fig.\,\ref{fig:var_rho_var_n_cher_scaling} for both components. The frequency is shifted up multiplying by $f_\rho$ for both components, the amplitude of $E_{\tilde{x}}$ (excess charge) is scaled up multiplying with $f_C=\sin\theta_{C_0}/\sin\theta_C$ and the amplitude of $E_y$ (geomagnetic effect) is scaled down, dividing by $f_\rho$. The patterns can be understood as a superposition of the two effects described before, the reduction of refractive index and the density. The result is similar to that of Fig.\,\ref{fig:var_rho_cte_n_ask} except for the fact that as the density drops there is an increase in the high frequency components because the decrease in density also implies a drop of the Cherenkov angle that reduces the destructive interference induced by the lateral spread of the shower. 

A dependence of the charge excess and geomagnetic current with air density has been reported in previous works \cite{Zilles:2018kwq, James:2022mea, Schluter:2022mhq, paudel2022simulation} where the relevance of the time delays has been identified. Here we have discussed in depth and check the scaling laws that apply. 
The complex scaling behavior of the radio pulses described above could be of particular importance to the "radio-morphing" approach~\cite{Zilles:2018kwq} that relies on modifications of pre-computed pulses using scaling properties for the showers. 

\subsection{Equivalence between changes in  magnetic field strength and air density}

It is particularly revealing to compare the effect of increasing the magnetic field strength to that of decreasing air density without changing the refractive index. We take the results for fixed magnetic field intensity $B_0=0.5\,\mathrm{G}$ and different density reduction factors $f_\rho=\rho_0/\rho$ with $\rho_0$ the sea level air density, and plot the spectral amplitudes of the electric field scaling the frequency with the factor $f_\rho$ (top right panel of Fig.\,\ref{fig:var_rho_cte_n_ask} and bottom right panel of Fig.\,\ref{fig:var_rho_cte_n_ask} without the scaling of the amplitude with $\rho^{-1}$). These are compared to simulations made for fixed air density $\rho_0$ and different magnetic field intensities, (left panels of Figs.\,\ref{fig:magnetic_field_strength_Askaryan} and \ref{fig:magnetic_field_strength_polarizations}). 

\begin{figure}[ht]
	\centering
	\subfigure{\includegraphics[width=0.49\linewidth]{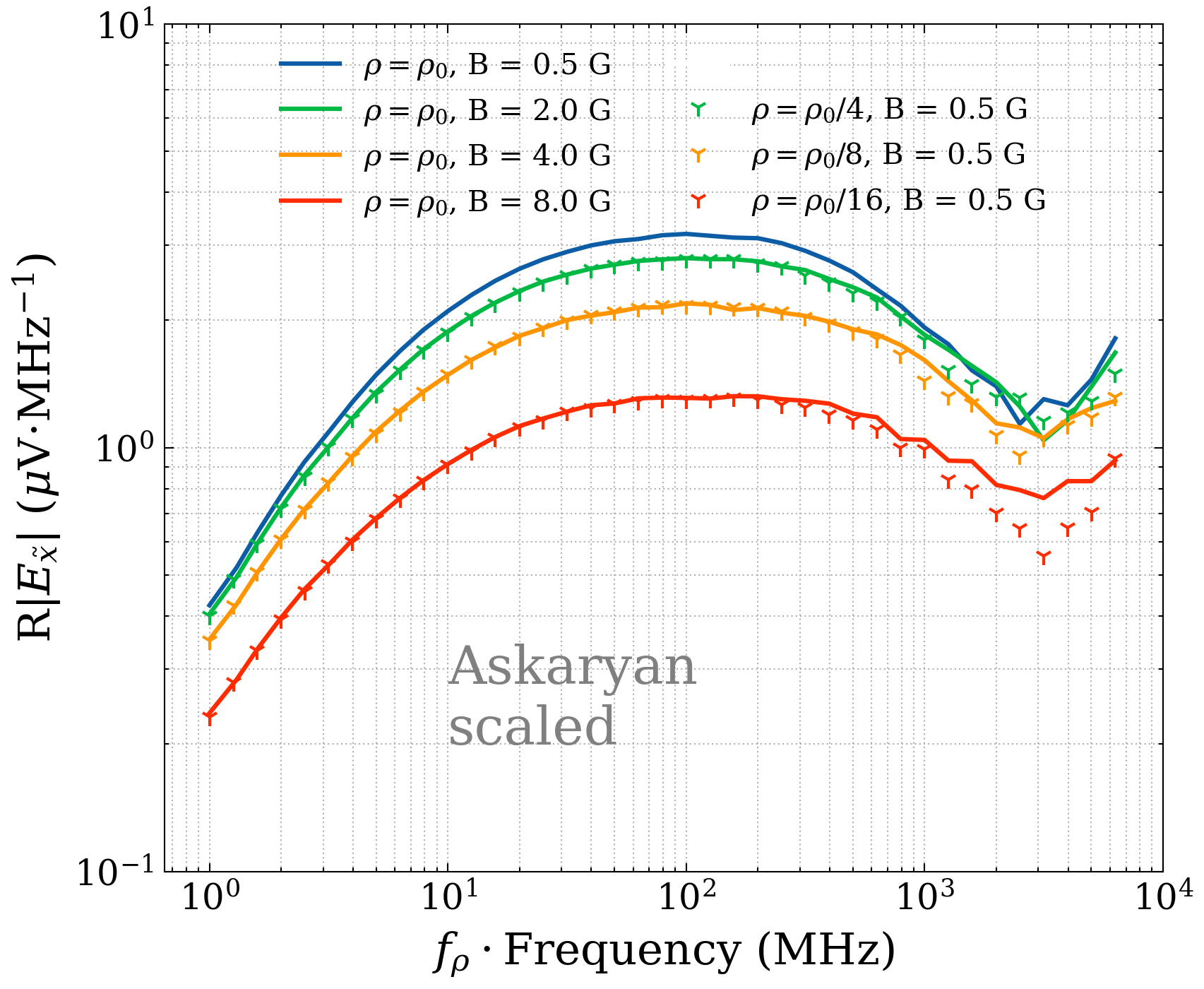}}
	\subfigure{\includegraphics[width=0.49\linewidth]{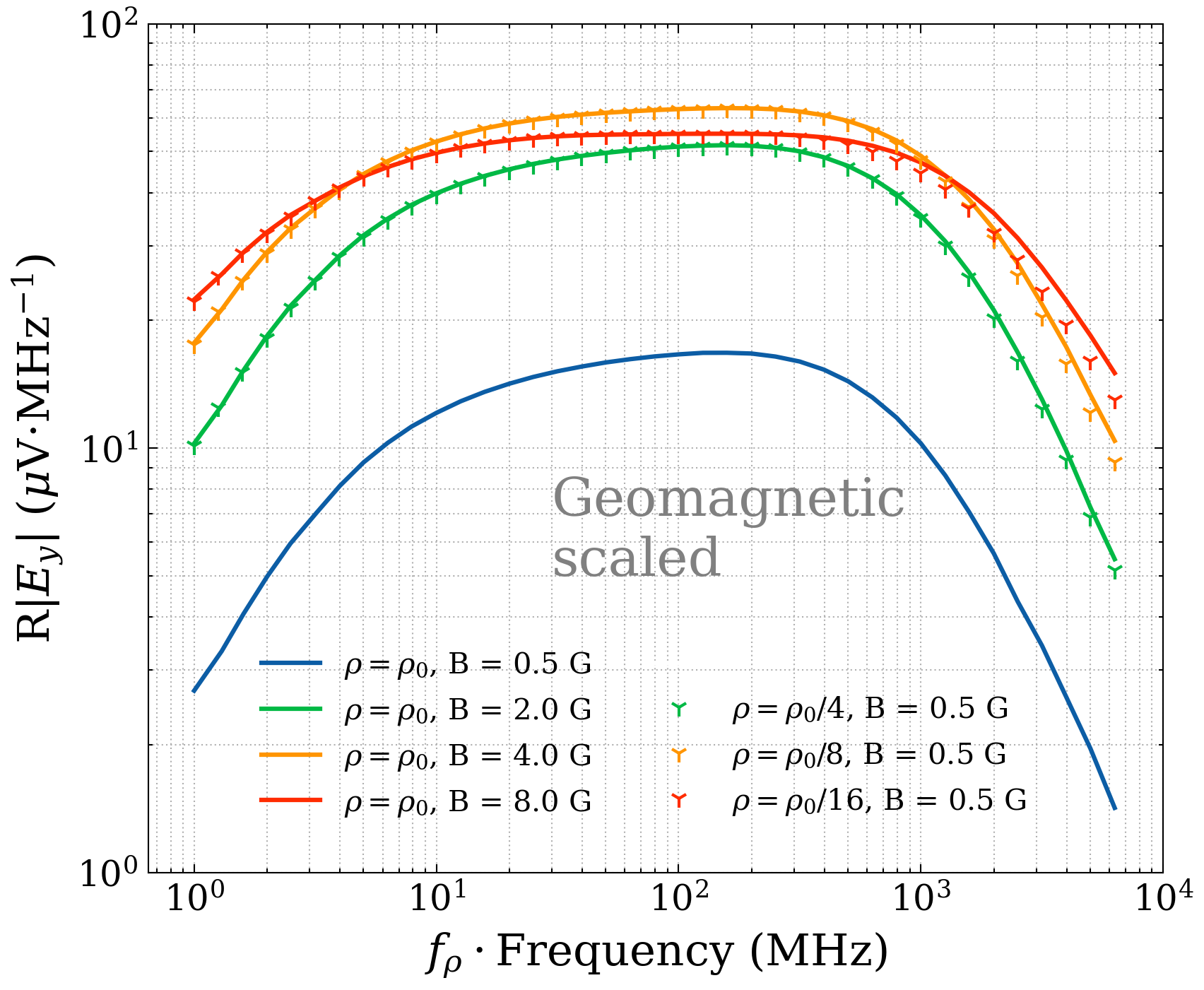}}
\caption{Average of the spectral amplitudes of the radio pulses of 50 showers initiated by a 100 TeV electron developing in air at different densities (with constant refractive index) and different magnetic fields using the same scaling factors. The observer is placed at the Cherenkov angle for the geometry shown in Fig.~\ref{fig:Geometry}. The $E_{\tilde{x}}$ components are plotted on the left and the $E_y$ components on the right. The frequency  multiplied by $f_\rho$ to account for the change of scale of the showers.}
\label{fig:magnetic_field_strength_comparison}
\end{figure}

The two families of curves are directly compared in  Fig.~\ref{fig:magnetic_field_strength_comparison}. The left and right panels respectively display the  polarization due to the excess charge ($E_{\tilde{x}}$) and to the geomagnetic effect ($E_y$). The level of agreement between the families of curves in the two cases is outstanding. Once we compensate the increase in the shower dimensions by scaling the frequencies with the $f_\rho$ factor the amplitudes have the same behavior: increasing the magnetic field is practically equivalent to reducing the density by the same factor without changing the refractive index (ignoring the effects that are due to the fact that the Cherenkov angle decreases in  lower density air). 
 What becomes relevant for interference is the radius of curvature of the particles relative to the shower dimensions, as has been already stated in \cite{James:2022mea}, where the time delays associated to the geomagnetic effect have been shown to increase with $B$. The radius of curvature is in turn proportional to $B^{-1}$ while the physical size of the shower is proportional to $\rho^{-1}$.

\subsection{Comparison to real case scenarios}

\begin{figure}[hbt]
\centering
\includegraphics[width=0.8\linewidth]{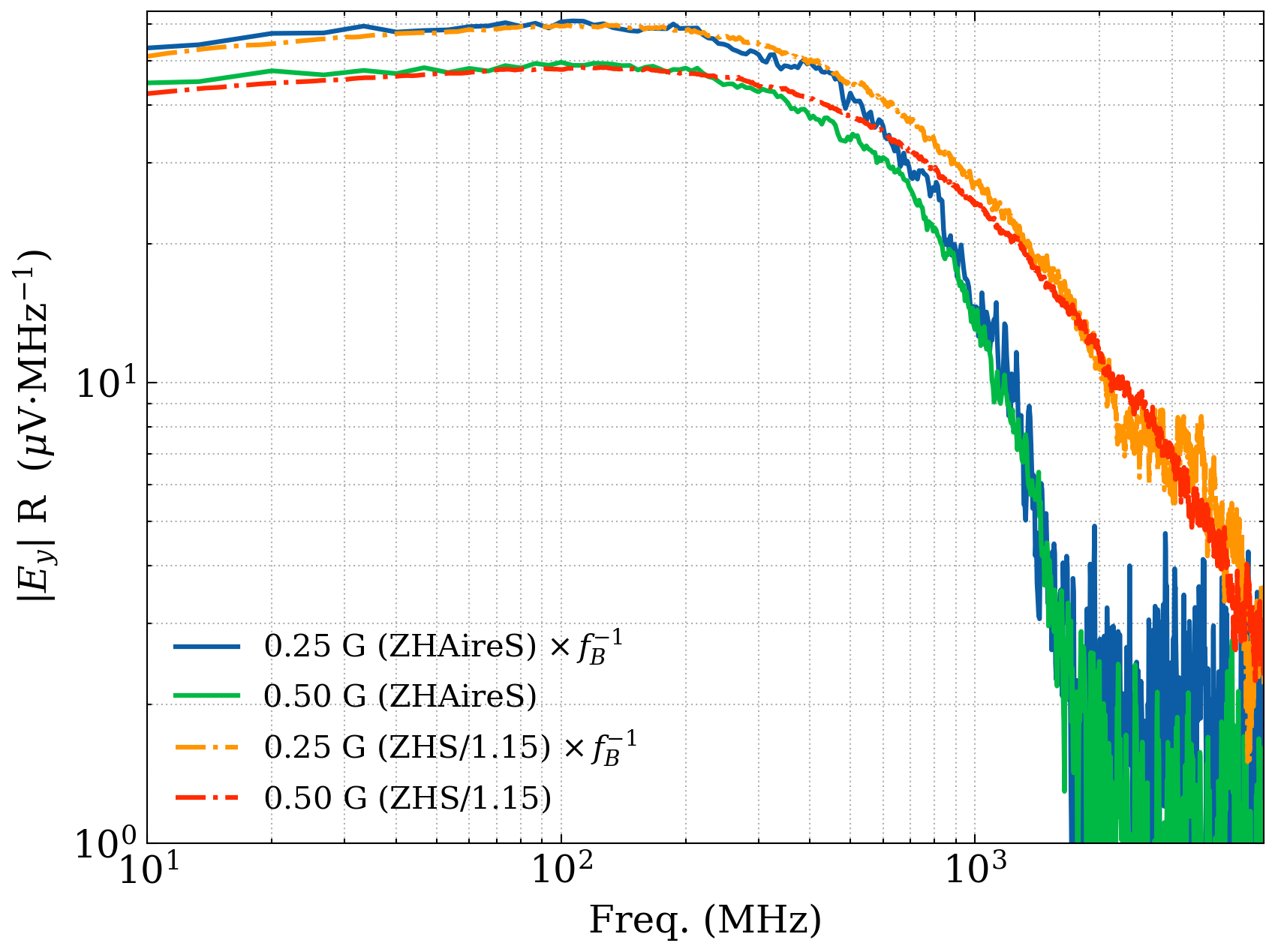}
\caption{Frequency spectra of the electric field amplitudes (multiplied by distance to the observer) for two showers simulated with ZHS in air of constant density (0.3\,mg\,cm$^{-3}$) compared to other two obtained with ZHAireS in a realistic atmosphere injected at 14.5 km above the ground with zenith angle $85^\circ$. All showers are initiated by 100 TeV electrons. Two cases with a magnetic field intensities of $B=0.25$\,G and 0.5\,G, both in the direction perpendicular to the shower axis, are displayed for each simulation program. The observer in ZHS is placed in the Cherenkov angle in the Fraunhofer regime for a geometry as sketched in Fig.\,\ref{fig:Geometry}. In ZHAireS simulations the observer is placed at the Earth's surface at (-20806, 0., -33.98) m coordinates in the ZHAireS reference system with origin at the shower impact point. This point approximately corresponds to the Cherenkov direction (calculated at the position of shower maximum). ZHAireS amplitudes have been multiplied by the distance to shower maximum and ZHS amplitudes have been divided by 1.15 to correct for a minor normalization mismatch (see text). Amplitudes for $B=0.25$\,G have been re-scaled with a factor of 2 (dividing by $f_B=0.5$). The plot displays the same violations of the linear scaling with $B$ at a level of $22\%$ when the magnetic field changes from 0.5 to 0.25~G in both simulations.}
\label{fig:realistic_inclined_85}
\end{figure}

Up to now we have been considering simplified simulations in a constant density air and with the observer in the Fraunhofer limit. These conditions can be rather far from reality and one can raise the question of to what extent these effects and scalings are applicable to more realistic showers in the atmosphere. We illustrate that some of these findings apply in realistic atmospheric showers by directly comparing an unthinned pulse of an electron shower of 100 TeV as obtained with ZHS in an atmosphere of constant density of 0.3~mg\,cm$^{-3}$, corresponding to sea level density reduced by a factor $f_\rho=4$, with a more realistic simulation using ZHAireS of a unthinned electron shower of 100 TeV incident in the atmosphere and impacting the Earth's surface with a zenith angle of $85^\circ$. The electron is injected at an altitude of $\sim 14.50\,\mathrm{km}$ leading to the occurrence of shower maximum at a depth in the atmosphere where the air density is close to the chosen value of 0.3~mg\,cm$^{-3}$. The observer is placed at sea level on the Earth's surface observing the shower in the direction of the Cerenkov angle corresponding to that density, at a distance of about 147 km from the position of shower maximum. In the ZHS simulation the observer is located at the Cherenkov angle in the Fraunhofer regime (see Table~\ref{tab:x_max_magnitudes}). 
The electric field spectra of these showers are shown in Fig.\,\ref{fig:realistic_inclined_85}, where we consider two cases of magnetic field strength $B=0.25$\,G and 0.5\,G, perpendicular to the shower axis. The amplitude of the ZHAireS simulations is multiplied by distance to shower maximum, $\sim 146.8 \times 10^{3}$~m, to compare to ZHS  normalization (obtained in the Fraunhofer limit) which is $15 \%$ higher. This has been corrected for reducing the ZHS amplitudes by $15\%$ to show that the pulses have the same shape below about 300 MHz~\footnote{This is not unexpected since the ZHS does not account for an exponential drop of density with altitude nor for changes in distance as the shower develops.}. 
The pulses obtained for $B=0.25$\,G have been multiplied by a factor of 2, so that they would coincide with those of $B=0.5$\,G in the ideal case of perfect scaling with $B$.

Fig.\,\ref{fig:realistic_inclined_85} illustrates that many conclusions obtained with simulated showers developing in constant density air are also relevant in more realistic conditions. The curves obtained for magnetic field intensities $B=0.25$\,G and 0.5\,G do not line up exactly, the latter being $\sim 22\%$ below those for $B=0.25$\,G. Deviations from linear scaling with $B$ are quoted in the frequency range $10-300$\,MHz, and they are due to magnetic deflections that reduce the number of particles contributing fully coherently as explained before. It is remarkable that the same drop of $\sim 22\%$ for the $B=0.5$\,G amplitudes relative to the 0.25\,G curves is obtained in both Monte Carlo simulations. The combination of density and magnetic field strength were chosen such that the showers were expected to display deviations from linear scaling, as illustrated in the previous subsection. This shows that scaling violations can even be observed at typical Earth magnetic field intensities and at low enough densities such as those at the position of shower maximum in very inclined showers \cite{Chiche:2021iin}. It can also be seen that the spectrum of the electric field amplitude of both simulations  agrees in shape up to about 700\,MHz, well above the bandwidths chosen for many experimental setups. Significant differences can be appreciated at higher frequencies where a faster decrease of the amplitude is seen in ZHAireS simulations. These may be attributed to the fact that in ZHAireS simulations the early part of the shower develops in very thin air and thus the time delays between the high-energy particles in the shower are larger than in ZHS simulations in which the density is constant.

\section{Summary and discussion}

We have developed an extension of the ZHS simulation program to account for magnetic deflections of charged particles. We have compared the longitudinal and lateral profiles of electron showers with this new code to those obtained with ZHAireS in air at constant density showing that they give compatible results at the $\sim5\%$ level (Fig.\,\ref{fig:aires_comparison}). 
Radio pulses have been simulated for showers developing in constant air density accounting for the geomagnetic effect for observers in the Cherenkov direction and in the Fraunhofer limit. The geomagnetic component of the pulses obtained with both programs are shown to give compatible results at a $\sim 5\%$ precision level (Fig.\,\ref{fig:pulse_comparison}). Some differences in the amplitudes for the Askaryan component have been observed which can be possibly attributed to a different treatment of energy loss and time delays. Nevertheless, the Askaryan component is typically subdominant relative to the geomagnetic and it is not expected to have too much of an impact in most relevant scenarios. 

The qualitative behavior of the frequency spectrum (Fig.\,\ref{fig:ice_radiopulse}), has been explained emphasizing the essential role of the time delays of particles in the shower. 
At sea level density, the spectral amplitude rises almost linearly with frequency up to $\sim 5$\,MHz. The time delays $t_D$ of the shower particles relative to a plane perpendicular to the shower direction moving at the speed of light are responsible for the loss of coherence of the particles in the Cherenkov direction for frequencies above $\sim 5$~MHz. As the frequency increases the pool of particles contributing fully coherently reduces, flattening the spectrum up to $\sim 200$\,MHz. Many experiments searching for cosmic-ray showers in air using the radio technique use bands within this flat region. At higher frequencies the time delays $t_r$, due to the lateral spread of the shower, further reduce the number of particles that contribute coherently explaining the drop with frequency seen in the spectrum.    

The new ZHS program has been used to study radio pulses in air in a comprehensive way, studying the effects of changing the magnetic field, the density of the air and the refractive index independently for observers in the Cherenkov direction. These comparisons have revealed complex scaling patterns of the radio pulses with deviations from a simple linear behaviour. The scaling laws obtained are different in the case of the two mechanisms at play, the charge excess (Askaryan effect), and the geomagnetic effect. These are separated in our simulations by adopting the geometry in Fig.\,\ref{fig:Geometry}. Several conclusions have been obtained from these studies:

\begin{itemize}

\item In air at sea level density as long as the magnetic field intensity $B<1$\,G, the amplitude of the 
geomagnetic component accurately scales with $B$, as often assumed, while the amplitude of the excess charge component has been shown to be largely independent of $B$ (Fig.\,\ref{fig:magnetic_field_strength_Askaryan}). 
For larger values of the magnetic field, deviations from the linear scaling of progressively increasing magnitude show up both for the excess charge and geomagnetic effects (Figs.\,\ref{fig:magnetic_field_strength_Askaryan} and \ref{fig:magnetic_field_strength_polarizations}). We have shown that this is due to the increasing importance of delays associated to the deflection of the particles in the magnetic field.

\item The effect of changing the density of air is twofold. A decrease of the density increases the shower dimensions, the particle tracklength 
 and the time delays, and reduces the Cherenkov angle all at the same time. The mean free path for particle interactions also increases, so that the transverse component of the current induced by the magnetic field has more time to freely develop without particle collisions. 

The frequency at which decoherence sets in is reduced in proportion to the decrease in density due to the larger shower dimensions and time delays. Simultaneously, the tracklength increases with the inverse density. As a consequence, the frequency spectra of the Askaryan component for different densities align themselves when artificially shifting up the frequency with the inverse density (top right panel in Fig.\,\ref{fig:var_rho_cte_n_ask}). However, at the highest frequencies the pulses from lower density air display more coherence because of the reduced Cherenkov angle. 

The geomagnetic effect behaves similarly but the amplitudes can be shown to have a further increase as the density drops. If in addition of the re-scaling of frequency, we plot the amplitudes of the geomagnetic effect scaled with the inverse of the air density, they line up for low densities. However, when the magnetic delays play a significant role, the spectral curves show increasing drops in amplitude relative to the expected linear behaviour (bottom right panel in  Fig.\,\ref{fig:var_rho_cte_n_ask}). This behaviour is attributed to the fact that the transverse currents that develop in reduced air density are stronger as they have more time to build up between particle interactions. But this also implies that the particles describe more closed arcs of a circle and therefore get further away from the axis accumulating more time delays. In this way, the increase in the transverse current as the density decreases is compensated (and even overcompensated) by the loss of coherence of a fraction of the particles that get out of phase because of the deflections in the magnetic field. 

\item Finally, we have shown that increasing the magnetic field has a remarkably similar effect to reducing the density when leaving the Cherenkov angle unchanged. For showers with reduced density the frequency must be scaled up with the density reduction factor $f_\rho$ to compensate for the change of dimensions of the shower. Once this correction is made the spectrum obtained with density $\rho$ and magnetic field intensity $B$ is equivalent to the spectrum obtained for density $\rho/f$ and intensity $B/f$ with $f$ an arbitrary constant (Fig.\,\ref{fig:magnetic_field_strength_comparison}). The violations of the scaling observed as the magnetic field increases above $\sim 1$\,G for sea level density are also observed as the density decreases. Naturally, as the density is reduced, the critical value of $B$ at which scaling deviations start taking place is also reduced.

\end{itemize}

The complex scaling with density and magnetic field obtained in the frequency domain has to be taken into account in the discussion of any scaling at the level of the peak amplitude or energy fluence as has been previously addressed in different experimental situations~\cite{Chiche:2021iin,PierreAuger:2016vya}. If the showers to be detected span a large range of densities, for instance for cosmic-ray showers of different zenith angles, the frequency shift required to compare different zenith angles must be explored in relation to the chosen frequency band. The frequency shift is bound to have an impact on the scaling properties of pulse amplitude and energy fluence.

To end, we have shown that the simulations made in homogeneous air are not only relevant to illustrate the complex behavior of electric field amplitudes as a function of $\rho$, $n$ and $B$. 
By comparing pulses in homogeneous air to pulses in an exponential atmosphere, it has been shown that the homogeneous air approximation can be useful to describe pulses produced by showers provided the density is chosen to be that corresponding to shower maximum (Fig.\,\ref{fig:magnetic_field_strength_comparison}). Moreover, the quantitative violation of linear scaling with density and magnetic field obtained in homogeneous air quite accurately reproduces that obtained in more realistic simulations at large zenith angles that account for the decrease of density with altitude.

\section{Acknowledgements}
We thank M. Tueros for comments after carefully reading the manuscript. This work has received financial support from 
Xunta de Galicia (Centro singular de investigación de Galicia accreditation 2019-2022); 
European Union ERDF;
''Mar\'\i a de Maeztu” Units of Excellence program MDM-2016-0692 and the Spanish Research State Agency,
Ministerio de Ciencia e Innovaci\'on/Agencia Estatal de Investigaci\'on PID2019-105544GB-I00;
RED2018-102661-T (RENATA); PRE2020-092276;
and Xunta de Galicia Consolidaci\'on 2021 GRC GI-2033 ED431C-2021/22 and Consolidaci\'on 2022 ED431F-2022/15.

\begin{appendices}
\section{Treatment of the Lorentz force}
\label{Appendix1}

\begin{figure}[ht]
\centering
\subfigure{\includegraphics[width=0.49\linewidth]{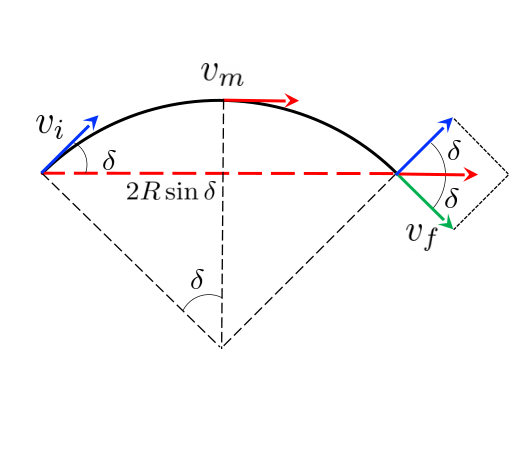}}
\subfigure{\includegraphics[width=0.49\linewidth]{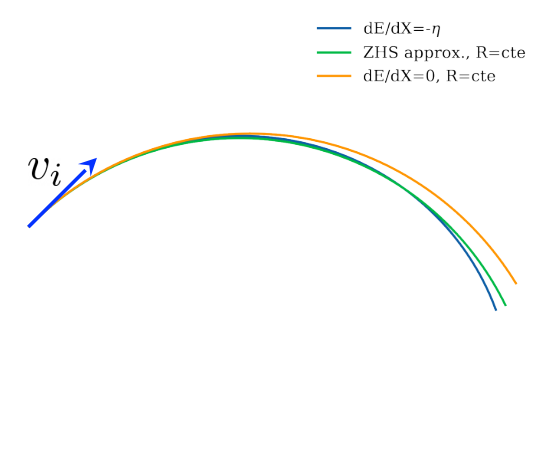}}    
\caption{Trajectories in the rotation plane, perpendicular to the magnetic field ${\bf B}$. Left: sketch of a trajectory with no energy loss which is an arc of a circle with a total deflection angle $2\delta$ and corresponding velocity changes. The red straight line shows the net displacement vector of length $2R\sin \delta$ in this plane in case of no multiple scattering. Right: Comparison of a trajectory in the rotation plane with no energy loss (orange) to the true trajectory for constant energy loss(blue) and to the arc of a circle used in ZHS. The scale is arbitrary. The radius is chosen so that that the path length coincides and the velocity has the same rotation as calculated analytically (see text). Large arcs have been plotted for clarity, typical arcs of particle trajectories have less rotation.
}
\label{fig:deflection_sketch}
\end{figure}

In the Monte Carlo simulation the propagation steps for a particle are calculated sampling the matter depth traveled before the next particle interaction, taking into account all the relevant interaction cross sections. These steps are further subdivided to make the treatment of continuous energy loss, multiple elastic scattering and magnetic deflections more accurate. In ZHS all distances are actually expressed in units of matter depth multiplying them by the density. Let us consider a given step of depth $\Delta X$. 
The charged particle trajectories under the influence of the Lorentz force are described separating the motion in two components, one is parallel to the magnetic field vector and hence not affected by it. The other component is contained in the ``rotation plane'' perpendicular to the magnetic field vector (left panel of  Fig.\,\ref{fig:deflection_sketch}) so that the particle describes a curve in this plane with radius of curvature $R$ given by: 
\begin{equation}
\frac{1}{R}=\frac{q B}{p\sin\alpha},
\label{eq:radius}
\end{equation}
where $B$ is the magnetic field strength, $q$ the charge of the particle, $p$ its momentum, and $\alpha$ the angle between the particle's velocity and the magnetic field. Consequently, the predefined matter depth travelled by the particle, $\Delta X$, is split in two components, one parallel to the magnetic field, $\Delta X_{\vert\vert} = \Delta X \cos \alpha$, and another that lies in the rotation plane, $\Delta X_\perp =\Delta X \sin \alpha$, corresponding to the two-dimensional projection of the trajectory onto that plane. 
Neglecting energy loss in the trajectory, it is relatively easy to calculate the rotation angle of the particle velocity in the rotation plane, $2 \delta$ (left panel of Fig.\,\ref{fig:deflection_sketch}). The particle describes an arc of a circle $2R\delta$ which, when expressed in g\,cm$^{-2}$, must coincide with $\Delta X_\perp =\Delta X\sin \alpha$. The displacement vector for the particle in the rotation plane has a direction ($\boldsymbol{v_m}$) which coincides with the sum of the start ($\boldsymbol{v_i}$) and end ($\boldsymbol{v_f}$) velocity vectors and its modulus is $2 R \sin(\delta)$.

Following the ZHS approach we have modified the calculation of $\delta$ taking into account continuous energy loss, recalculated the displacement vector considering the changing curvature along the arc as energy is lost, and finally made a correction to the displacement vector to account for Moli\`ere scattering. 
As a particle of energy $E$ loses energy along the track, the radius of curvature decreases. The particle loses energy in proportion to the matter depth of the  trajectory and we can write the energy loss per unit depth as:
\begin{equation}
    \frac{\dd E}{\dd X}=-\eta,
\label{eq:loss_rate}
\end{equation}
where $\eta$ is approximately constant over a large energy range. We also assume $\eta$ remains constant within each track and this allows analytic integration. In the rotation plane the infinitesimal rotation corresponding to an infinitesimal displacement of a circular trajectory is given by:
\begin{equation}
\dd\delta = \frac{\dd X_\perp}{R}
\label{eq:deflection}
\end{equation}
where $X_\perp = X \sin \alpha$ is used because this corresponds to the projection of the track onto the rotation plane. Using the relation between the radius and the energy 
given by Eq.~(\ref{eq:radius}) we integrate between initial, $E_0$, and final, $E_1$, energies to find the rotation angle for the particle velocity:
\begin{equation}
  \delta = \frac{qBc}{\eta} \log\frac{p_0 c + E_0}{p_1 c + E_1},
\label{eq:delta}
\end{equation}
which is independent of $\alpha$. Eq.\,(\ref{eq:delta}) is exact for a constant energy loss rate, but an approximation is still needed to calculate the displacement vector in the rotation plane for such a trajectory. We approximate it as if it was that of a perfect arc of a circle of the same length and a radius that produces the same velocity rotation. The Monte Carlo steps have been further subdivided to ensure that the fractional energy loss is always below $10\%$ of the initial energy for improved accuracy. 
At this stage we just need to add the displacement component in the rotation plane and that parallel to the magnetic field direction. 
We have checked the program approximates the expected curved trajectories testing it for individual tracks. The approximated tracks give a good description of the trajectory obtained with a very fine subdivision of the steps in which the radius of curvature is updated after having calculated the energy loss. 

Finally the effects of multiple scattering along the curved trajectory must be also accounted for. In the original ZHS program multiple scattering is taken into account by changing the direction of the particle after each step. Provided the angular deviations are small in Moli\`ere's treatment, the mean deflection angle of the particle velocity after traversing a depth $X$ is characterized by the polar angle relative to the original particle direction, which is approximated by~\cite{moliere1948theorie}:
\begin{equation}
    \Theta_{\rm MS} \simeq \frac{E_{\rm MS}}{E}\sqrt{\frac{X}{X_0}},
    \label{eq:ms_angle}
\end{equation}
where $X_0$ is the radiation length of the medium and $E_{\rm MS}=m_e\sqrt{{4\pi}/{\alpha_{\rm em}}}\simeq 21 \textrm{ MeV}$, 
$\alpha_{\rm em}$ being the fine structure constant. 
To define the full rotation of the initial velocity an  "azimuth" angle (measured in the transverse plane to the initial particle direction) must be specified. This angle is random so that the particle velocity can be treated as a random walk in the two orthogonal directions to the initial particle direction. As a result of multiple scattering the displacement vector in the transverse plane will have a small deviation relative to the non-scattered trajectory in a direction close to that defined by this  azimuth angle. In addition, the component of the displacement vector along the original particle direction will be slightly reduced. This is approximately accounted for splitting the rectilinear track in two equal halves, the first half keeps the original direction and the second half is  rotated about its start point an angle $\theta_{\rm MS}$, so the final velocity has the desired direction. 

\begin{figure}[hbt]
\centering
\includegraphics[width=0.7\linewidth]{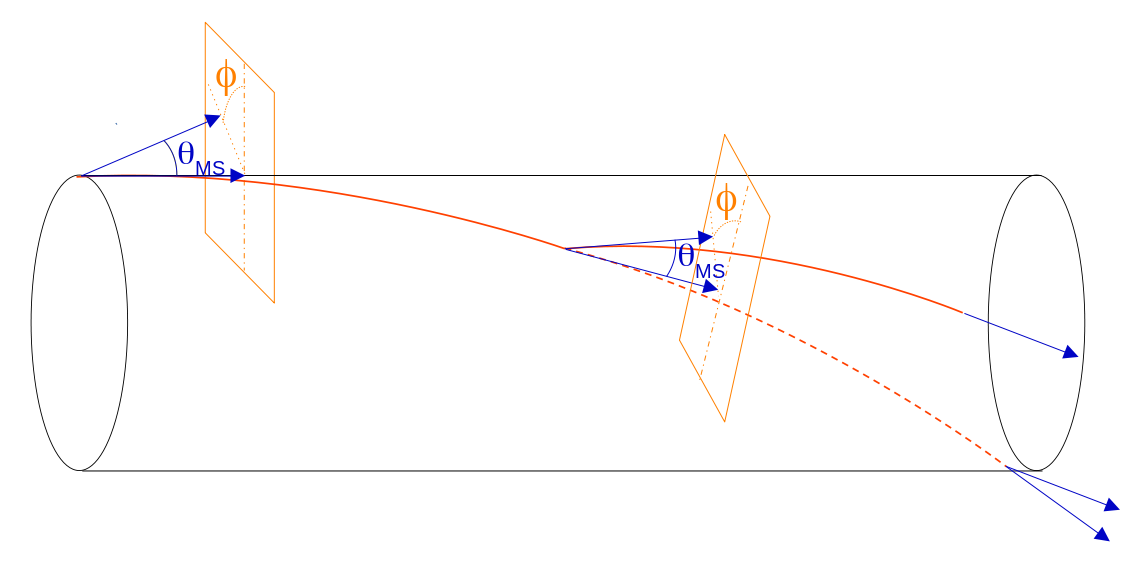}
\caption{Sketch of the modification of the trajectory followed by a charged particle in the presence of a magnetic field because of multiple elastic scattering as implemented in the new ZHS code. The scattering angle for the velocity is calculated as for a rectilinear trajectory. The approximated helix calculated for the magnetic deflection is split in two halves. The first half is left unmodified. The second half is rotated by the Moli\`ere angle, $\Theta_{\rm MS}$ and the azimuth angle, $\phi$, about the central point of the track, so that the final velocity vector coincides with the final velocity of the unmodified helical motion (dashed line) after applying the same rotation. 
}
\label{fig:multiple_scattering} 
\end{figure}

We have tried to maintain an equivalent approach for the treatment of curved trajectories. We first calculate the trajectory assuming that the magnetic deviation is unaffected by multiple elastic scattering. We then calculate the deviation from the original particle direction due to Moli\`ere scattering for the case of no magnetic deviation. We obtain the rotation matrix for the velocity vector as described above, through an angle $\Theta_{MS}$ as given by Eq.~(\ref{eq:ms_angle}) relative to the particle direction and with a random azimuth direction in the orthogonal plane as sketched at the start of the trajectory in Fig.~\ref{fig:multiple_scattering}. This deviation must be superimposed to the trajectory affected with the magnetic deflection. As in the case of no magnetic deflection, we split the curved track in two halves, leaving the first half unaffected by multiple scattering and applying the same rotation matrix to the second half of the trajectory about its start point. This procedure will provide a final particle direction consistent with the calculated one and will approximately account both for the extra time delay and the spatial shift. 

\end{appendices}

\bibliography{main}

\providecommand{\noopsort}[1]{}\providecommand{\singleletter}[1]{#1}%

\providecommand{\href}[2]{#2}\begingroup\raggedright\begin{thebibliography}{10}

\bibitem{JELLEY1965}
J.~V. Jelley, J.~H. Fruin, N.~A. Porter, T.~C. Weekes, F.~G. Smith and R.~A.
  Porter, \emph{Radio pulses from extensive cosmic-ray air showers},
  \href{https://doi.org/10.1038/205327a0}{\emph{Nature} {\bfseries 205} (1965)
  327}.

\bibitem{Gusev:1983wuo}
G.~A. Gusev and I.~M. Zheleznykh, \emph{{Neutrino and Muon Detection from the
  Radio Emission of Cascades created by them in Natural Dielectric Media}},
  {\emph{JETP Lett.} {\bfseries 38} (1983) 611}.

\bibitem{Miocinovic2005}
P.~Miocinovic, S.~W. Barwick, J.~J. Beatty, D.~Z. Besson, W.~R. Binns, B.~Cai
  et~al., \emph{Tuning into uhe neutrinos in antarctica - the anita
  experiment}, \href{https://doi.org/10.48550/
  arXiv:astro-ph/0503304}{\emph{ECONF C041213:2516,2004} (2005) }
  [\href{https://arxiv.org/abs/astro-ph/0503304}{{\ttfamily
  astro-ph/0503304}}].

\bibitem{Rottgering2003}
H.~Rottgering, \emph{Lofar, a new low frequency radio telescope},
  \href{https://doi.org/10.1016/S1387-6473(03)00057-5}{\emph{New Astron. Rev.}
  {\bfseries 47} (2003) 405}
  [\href{https://arxiv.org/abs/astro-ph/0309537}{{\ttfamily
  astro-ph/0309537}}].

\bibitem{Fuchs2012}
{\scshape Pierre Auger} collaboration, \emph{The auger engineering radio
  array}, \href{https://doi.org/10.1016/j.nima.2012.01.058}{\emph{Nucl.
  Instrum. Meth. A} {\bfseries 692} (2012) 93}.

\bibitem{Bezyazeekov2015}
P.~A. Bezyazeekov et~al., \emph{Measurement of cosmic-ray air showers with the
  tunka radio extension (tunka-rex)},
  \href{https://doi.org/10.1016/j.nima.2015.08.061}{\emph{Nucl. Instrum. Meth.
  A} {\bfseries 802} (2015) 89}
  [\href{https://arxiv.org/abs/1509.08624}{{\ttfamily 1509.08624}}].

\bibitem{zas1992electromagnetic}
E.~Zas, F.~Halzen and T.~Stanev, \emph{Electromagnetic pulses from high-energy
  showers: Implications for neutrino detection},
  \href{https://doi.org/https://doi.org/10.1103/PhysRevD.45.362}{\emph{Physical
  Review D} {\bfseries 45} (1992) 362}.

\bibitem{Alvarez-Muniz:2010wjm}
J.~Alvarez-Mu\~niz, A.~Romero-Wolf and E.~Zas, \emph{{Cherenkov radio pulses
  from electromagnetic showers in the time-domain}},
  \href{https://doi.org/10.1103/PhysRevD.81.123009}{\emph{Phys. Rev. D}
  {\bfseries 81} (2010) 123009}
  [\href{https://arxiv.org/abs/1002.3873}{{\ttfamily 1002.3873}}].

\bibitem{Alvarez-Muniz:2022uey}
J.~Alvarez-Mu\~niz and E.~Zas, \emph{{Progress in the Simulation and Modelling
  of Coherent Radio Pulses from Ultra High-Energy Cosmic Particles}},
  \href{https://doi.org/10.3390/universe8060297}{\emph{Universe} {\bfseries 8}
  (2022) 297}.

\bibitem{James:2010vm}
C.~W. James, H.~Falcke, T.~Huege and M.~Ludwig, \emph{{General description of
  electromagnetic radiation processes based on instantaneous charge
  acceleration in `endpoints'}},
  \href{https://doi.org/10.1103/PhysRevE.84.056602}{\emph{Phys. Rev. E}
  {\bfseries 84} (2011) 056602}
  [\href{https://arxiv.org/abs/1007.4146}{{\ttfamily 1007.4146}}].

\bibitem{AlvarezMuniz2012}
J.~Alvarez-Muniz, W.~R. Carvalho, Jr. and E.~Zas, \emph{Monte carlo simulations
  of radio pulses in atmospheric showers using zhaires},
  \href{https://doi.org/10.1016/j.astropartphys.2011.10.005}{\emph{Astropart.
  Phys.} {\bfseries 35} (2012) 325}
  [\href{https://arxiv.org/abs/1107.1189}{{\ttfamily 1107.1189}}].

\bibitem{Huege:2013vt}
T.~Huege, M.~Ludwig and C.~W. James, \emph{Simulating radio emission from air
  showers with coreas}, \href{https://doi.org/10.1063/1.4807534}{\emph{AIP
  Conf. Proc.} {\bfseries 1535} (2013) 128}
  [\href{https://arxiv.org/abs/1301.2132}{{\ttfamily 1301.2132}}].

\bibitem{kahn1966radiation}
F.~D. Kahn and I.~Lerche, \emph{Radiation from cosmic ray air showers},
  {\emph{Proceedings of the Royal Society of London. Series A. Mathematical and
  Physical Sciences} {\bfseries 289} (1966) 206}.

\bibitem{Clay:1969}
H.~R. Allan, R.~W. Clay, J.~K. Jones, A.~T. Abrosimov and K.~P. ~, \emph{{Radio
  Pulse Production in Extensive Air Showers}}, {\emph{Nature} {\bfseries 222}
  (1969) 635}.

\bibitem{James:2022mea}
C.~W. James, \emph{{Nature of radio-wave radiation from particle cascades}},
  \href{https://doi.org/10.1103/PhysRevD.105.023014}{\emph{Phys. Rev. D}
  {\bfseries 105} (2022) 023014}
  [\href{https://arxiv.org/abs/2201.01298}{{\ttfamily 2201.01298}}].

\bibitem{Huege:2003up}
T.~Huege and H.~Falcke, \emph{{Radio emission from cosmic ray air showers:
  Coherent geosynchrotron radiation}},
  \href{https://doi.org/10.1051/0004-6361:20031422}{\emph{Astron. Astrophys.}
  {\bfseries 412} (2003) 19}
  [\href{https://arxiv.org/abs/astro-ph/0309622}{{\ttfamily
  astro-ph/0309622}}].

\bibitem{chicheARENA2022}
S.~Chiche, \emph{New features in the radio-emission of very inclined showers
  (to be published)}, {\emph{ARENA} (2022) }.

\bibitem{Schroder:2016hrv}
F.~G. Schr\"oder, \emph{{Radio detection of Cosmic-Ray Air Showers and
  High-Energy Neutrinos}},
  \href{https://doi.org/10.1016/j.ppnp.2016.12.002}{\emph{Prog. Part. Nucl.
  Phys.} {\bfseries 93} (2017) 1}
  [\href{https://arxiv.org/abs/1607.08781}{{\ttfamily 1607.08781}}].

\bibitem{Connolly:2016pqr}
A.~L. Connolly and A.~G. Vieregg, \emph{{Radio Detection of High Energy
  Neutrinos}}, pp.~217--240.
\newblock World Scientific, mar, 2017.
\newblock \href{https://arxiv.org/abs/1607.08232}{{\ttfamily 1607.08232}}.
\newblock 10.1142/9789814759410\_0015.

\bibitem{Huege:2017khw}
T.~Huege and D.~Besson, \emph{{Radio-wave detection of ultra-high-energy
  neutrinos and cosmic rays}},
  \href{https://doi.org/10.1093/ptep/ptx009}{\emph{PTEP} {\bfseries 2017}
  (2017) 12A106} [\href{https://arxiv.org/abs/1701.02987}{{\ttfamily
  1701.02987}}].

\bibitem{Buniy:2000kk}
R.~V. Buniy and J.~P. Ralston, \emph{{Radio detection of high-energy particles:
  Coherence versus multiple scales}},
  \href{https://doi.org/10.1103/PhysRevD.65.016003}{\emph{Phys. Rev. D}
  {\bfseries 65} (2002) 016003}
  [\href{https://arxiv.org/abs/astro-ph/0003408}{{\ttfamily
  astro-ph/0003408}}].

\bibitem{Scholten:2007ky}
O.~Scholten, K.~Werner and F.~Rusydi, \emph{{A Macroscopic Description of
  Coherent Geo-Magnetic Radiation from Cosmic Ray Air Showers}},
  \href{https://doi.org/10.1016/j.astropartphys.2007.11.012}{\emph{Astropart.
  Phys.} {\bfseries 29} (2008) 94}
  [\href{https://arxiv.org/abs/0709.2872}{{\ttfamily 0709.2872}}].

\bibitem{Werner:2007kh}
K.~Werner and O.~Scholten, \emph{{Macroscopic Treatment of Radio Emission from
  Cosmic Ray Air Showers based on Shower Simulations}},
  \href{https://doi.org/10.1016/j.astropartphys.2008.04.004}{\emph{Astropart.
  Phys.} {\bfseries 29} (2008) 393}
  [\href{https://arxiv.org/abs/0712.2517}{{\ttfamily 0712.2517}}].

\bibitem{Alvarez-Muniz:2011wcg}
J.~Alvarez-Mu\~niz, A.~Romero-Wolf and E.~Zas, \emph{{Practical and accurate
  calculations of Askaryan radiation}},
  \href{https://doi.org/10.1103/PhysRevD.84.103003}{\emph{Phys. Rev. D}
  {\bfseries 84} (2011) 103003}
  [\href{https://arxiv.org/abs/1106.6283}{{\ttfamily 1106.6283}}].

\bibitem{Werner:2012cr}
K.~Werner, K.~D. de~Vries and O.~Scholten, \emph{{A Realistic Treatment of
  Geomagnetic Cherenkov Radiation from Cosmic Ray Air Showers}},
  \href{https://doi.org/10.1016/j.astropartphys.2012.07.007}{\emph{Astropart.
  Phys.} {\bfseries 37} (2012) 5}
  [\href{https://arxiv.org/abs/1201.4471}{{\ttfamily 1201.4471}}].

\bibitem{Alvarez-Muniz:2020ary}
J.~Alvarez-Mu\~niz, P.~M. Hansen, A.~Romero-Wolf and E.~Zas, \emph{{Askaryan
  radiation from neutrino-induced showers in ice}},
  \href{https://doi.org/10.1103/PhysRevD.101.083005}{\emph{Phys. Rev. D}
  {\bfseries 101} (2020) 083005}
  [\href{https://arxiv.org/abs/2003.09705}{{\ttfamily 2003.09705}}].

\bibitem{Zilles:2018kwq}
A.~Zilles, O.~Martineau-Huynh, K.~Kotera, M.~Tueros, K.~de~Vries,
  W.~Carvalho~Jr. et~al., \emph{{Radio Morphing: towards a fast computation of
  the radio signal from air showers}},
  \href{https://doi.org/10.1016/j.astropartphys.2019.06.001}{\emph{Astropart.
  Phys.} {\bfseries 114} (2020) 10}
  [\href{https://arxiv.org/abs/1811.01750}{{\ttfamily 1811.01750}}].

\bibitem{Tueros:2020buc}
M.~Tueros and A.~Zilles, \emph{{Synthesis of radio signals from extensive air
  showers using previously computed microscopic simulations}},
  \href{https://doi.org/10.1088/1748-0221/16/02/P02031}{\emph{JINST} {\bfseries
  16} (2021) P02031} [\href{https://arxiv.org/abs/2008.06454}{{\ttfamily
  2008.06454}}].

\bibitem{Alvarez-Muniz:2014wna}
J.~Alvarez-Mu\~niz, W.~R. Carvalho, Jr., H.~Schoorlemmer and E.~Zas,
  \emph{{Radio pulses from ultra-high energy atmospheric showers as the
  superposition of Askaryan and geomagnetic mechanisms}},
  \href{https://doi.org/10.1016/j.astropartphys.2014.04.004}{\emph{Astropart.
  Phys.} {\bfseries 59} (2014) 29}
  [\href{https://arxiv.org/abs/1402.3504}{{\ttfamily 1402.3504}}].

\bibitem{Glaser:2018byo}
C.~Glaser, S.~de~Jong, M.~Erdmann and J.~R. H\"orandel, \emph{{An analytic
  description of the radio emission of air showers based on its emission
  mechanisms}},
  \href{https://doi.org/10.1016/j.astropartphys.2018.08.004}{\emph{Astropart.
  Phys.} {\bfseries 104} (2019) 64}
  [\href{https://arxiv.org/abs/1806.03620}{{\ttfamily 1806.03620}}].

\bibitem{Schluter:2022mhq}
F.~Schl\"uter and T.~Huege, \emph{{Signal model and event reconstruction for
  the radio detection of inclined air showers}},
  \href{https://doi.org/10.1088/1475-7516/2023/01/008}{\emph{JCAP} {\bfseries
  01} (2023) 008} [\href{https://arxiv.org/abs/2203.04364}{{\ttfamily
  2203.04364}}].

\bibitem{PierreAuger:2016vya}
{\scshape Pierre Auger} collaboration, \emph{{Measurement of the Radiation
  Energy in the Radio Signal of Extensive Air Showers as a Universal Estimator
  of Cosmic-Ray Energy}},
  \href{https://doi.org/10.1103/PhysRevLett.116.241101}{\emph{Phys. Rev. Lett.}
  {\bfseries 116} (2016) 241101}
  [\href{https://arxiv.org/abs/1605.02564}{{\ttfamily 1605.02564}}].

\bibitem{falcke2005detection}
H.~Falcke, W.~Apel, A.~Badea, L.~B{\"a}hren, K.~Bekk, A.~Bercuci et~al.,
  \emph{Detection and imaging of atmospheric radio flashes from cosmic ray air
  showers}, {\emph{Nature} {\bfseries 435} (2005) 313}.

\bibitem{aab2014probing}
A.~Aab, P.~Abreu, M.~Aglietta, M.~Ahlers, E.~Ahn, I.~Albuquerque et~al.,
  \emph{Probing the radio emission from air showers with polarization
  measurements}, {\emph{Physical Review D} {\bfseries 89} (2014) 052002}.

\bibitem{ardouin2009geomagnetic}
D.~Ardouin, A.~Belletoile, C.~Berat, D.~Breton, D.~Charrier, J.~Chauvin et~al.,
  \emph{Geomagnetic origin of the radio emission from cosmic ray induced air
  showers observed by codalema}, {\emph{Astroparticle Physics} {\bfseries 31}
  (2009) 192}.

\bibitem{T-510:2015pyu}
{\scshape T-510} collaboration, \emph{{Accelerator measurements of
  magnetically-induced radio emission from particle cascades with applications
  to cosmic-ray air showers}},
  \href{https://doi.org/10.1103/PhysRevLett.116.141103}{\emph{Phys. Rev. Lett.}
  {\bfseries 116} (2016) 141103}
  [\href{https://arxiv.org/abs/1507.07296}{{\ttfamily 1507.07296}}].

\bibitem{deVries:2011pa}
K.~D. de~Vries, A.~M. van~den Berg, O.~Scholten and K.~Werner, \emph{{Coherent
  Cherenkov Radiation from Cosmic-Ray-Induced Air Showers}},
  \href{https://doi.org/10.1103/PhysRevLett.107.061101}{\emph{Phys. Rev. Lett.}
  {\bfseries 107} (2011) 061101}
  [\href{https://arxiv.org/abs/1107.0665}{{\ttfamily 1107.0665}}].

\bibitem{Romero-Wolf:2021D+}
A.~Romero-Wolf, \emph{{Radio Simulations of Upgoing Extensive Air Showers
  Observed from Low-Earth Orbit}},
  \href{https://doi.org/10.22323/1.395.1031}{\emph{PoS} {\bfseries ICRC2021}
  (2021) 1031}.

\bibitem{Chiche:2021iin}
S.~Chiche, O.~Martineau-Huynh, K.~Kotera, M.~Tueros and K.~D.~de Vries,
  \emph{{Radio-Morphing: a fast, efficient and accurate tool to compute the
  radio signals from air-showers}},
  \href{https://doi.org/10.22323/1.395.0194}{\emph{PoS} {\bfseries ICRC2021}
  (2021) 194} [\href{https://arxiv.org/abs/2202.05886}{{\ttfamily
  2202.05886}}].

\bibitem{AlvarezMuniz2009}
J.~Alvarez-Muniz, C.~W. James, R.~J. Protheroe and E.~Zas, \emph{Thinned
  simulations of extremely energetic showers in dense media for radio
  applications},
  \href{https://doi.org/10.1016/j.astropartphys.2009.06.005}{\emph{Astropart.
  Phys.} {\bfseries 32} (2009) 100}.

\bibitem{AlvarezMuniz2012a}
J.~Alvarez-Mu{\~n}iz, W.~R. Carvalho, M.~Tueros and E.~Zas, \emph{Coherent
  cherenkov radio pulses from hadronic showers up to eev energies},
  \href{https://doi.org/10.1016/j.astropartphys.2011.10.002}{\emph{Astroparticle
  Physics} {\bfseries 35} (2012) 287}
  [\href{https://arxiv.org/abs/1005.0552}{{\ttfamily 1005.0552}}].

\bibitem{Motloch:2015wca}
P.~Motloch, J.~Alvarez-Mu\~niz, P.~Privitera and E.~Zas, \emph{{Transition
  radiation at radio frequencies from ultrahigh-energy neutrino-induced
  showers}}, \href{https://doi.org/10.1103/PhysRevD.93.043010}{\emph{Phys. Rev.
  D} {\bfseries 93} (2016) 043010}
  [\href{https://arxiv.org/abs/1509.01584}{{\ttfamily 1509.01584}}].

\bibitem{Bechtol:2021tyd}
K.~Bechtol et~al., \emph{{SLAC T-510 experiment for radio emission from
  particle showers: Detailed simulation study and interpretation}},
  \href{https://doi.org/10.1103/PhysRevD.105.063025}{\emph{Phys. Rev. D}
  {\bfseries 105} (2022) 063025}
  [\href{https://arxiv.org/abs/2111.04334}{{\ttfamily 2111.04334}}].

\bibitem{Askaryan:1962}
G.~A. Askar'yan, \emph{{Excess Negative Charge of an Electron-Photon Shower and
  its Coherent Radio Emission}}, {\emph{Soviet Physics JETP} {\bfseries 14}
  (1962) 441}.

\bibitem{Alvarez:1995}
J.~Alvarez-Muniz, G.~Parente and E.~Zas, \emph{{Radio Detection of High Energy
  Showers}},  in \emph{{24th International Cosmic Ray Conference}}, vol.~1,
  p.~1023, 8, 1995.

\bibitem{Alvarez-Muniz:2000aah}
J.~Alvarez-Muniz, R.~A. Vazquez and E.~Zas, \emph{{Calculation methods for
  radio pulses from high-energy showers}},
  \href{https://doi.org/10.1103/PhysRevD.62.063001}{\emph{Phys. Rev. D}
  {\bfseries 62} (2000) 063001}
  [\href{https://arxiv.org/abs/astro-ph/0003315}{{\ttfamily
  astro-ph/0003315}}].

\bibitem{Sciutto:1999jh}
S.~J. Sciutto, \emph{{AIRES: A system for air shower simulations}},
  \href{https://arxiv.org/abs/astro-ph/9911331}{{\ttfamily astro-ph/9911331}}.

\bibitem{AIRES_19_04_00}
S.~Sciutto, ``Aires user's manual and reference guide; version 19.04.00.''
  \url{http://aires.fisica.unlp.edu.ar/}, 2019.

\bibitem{CORSIKA}
D.~Heck, J.~Knapp, J.~Capdevielle, G.~Schatz and T.~Thouw, ``Corsika: A monte
  carlo code to simulate extensive air showers.'' FZKA Report 6019,
  Forschungszentrum Karlsruhe, 1998.

\bibitem{Garcia-Fernandez:2012urf}
D.~Garcia-Fernandez, J.~Alvarez-Muniz, W.~R. Carvalho, A.~Romero-Wolf and
  E.~Zas, \emph{{Calculations of electric fields for radio detection of
  ultrahigh energy particles}},
  \href{https://doi.org/10.1103/PhysRevD.87.023003}{\emph{Phys. Rev. D}
  {\bfseries 87} (2013) 023003}
  [\href{https://arxiv.org/abs/1210.1052}{{\ttfamily 1210.1052}}].

\bibitem{cazon2004time}
L.~Cazon, R.~Vazquez, A.~Watson and E.~Zas, \emph{Time structure of muonic
  showers}, {\emph{Astroparticle Physics} {\bfseries 21} (2004) 71}.

\bibitem{alvarez2012coherent}
J.~Alvarez-Muniz, W.~R. Carvalho~Jr, A.~Romero-Wolf, M.~Tueros and E.~Zas,
  \emph{Coherent radiation from extensive air showers in the ultrahigh
  frequency band}, {\emph{Physical Review D} {\bfseries 86} (2012) 123007}.

\bibitem{Alvarez-Muniz:2010hbb}
J.~Alvarez-Muniz, W.~R. Carvalho, Jr., M.~Tueros and E.~Zas, \emph{{Coherent
  Cherenkov radio pulses from hadronic showers up to EeV energies}},
  \href{https://doi.org/10.1016/j.astropartphys.2011.10.002}{\emph{Astropart.
  Phys.} {\bfseries 35} (2012) 287}
  [\href{https://arxiv.org/abs/1005.0552}{{\ttfamily 1005.0552}}].

\bibitem{nigl2008frequency}
A.~Nigl, W.~Apel, J.~Arteaga, T.~Asch, J.~Auffenberg, F.~Badea et~al.,
  \emph{Frequency spectra of cosmic ray air shower radio emission measured with
  lopes}, {\emph{Astronomy \& Astrophysics} {\bfseries 488} (2008) 807}.

\bibitem{ANITA:2010ect}
{\scshape ANITA} collaboration, \emph{{Observation of Ultra-high-energy Cosmic
  Rays with the ANITA Balloon-borne Radio Interferometer}},
  \href{https://doi.org/10.1103/PhysRevLett.105.151101}{\emph{Phys. Rev. Lett.}
  {\bfseries 105} (2010) 151101}
  [\href{https://arxiv.org/abs/1005.0035}{{\ttfamily 1005.0035}}].

\bibitem{ANITA:2018sgj}
{\scshape ANITA} collaboration, \emph{{Observation of an Unusual Upward-going
  Cosmic-ray-like Event in the Third Flight of ANITA}},
  \href{https://doi.org/10.1103/PhysRevLett.121.161102}{\emph{Phys. Rev. Lett.}
  {\bfseries 121} (2018) 161102}
  [\href{https://arxiv.org/abs/1803.05088}{{\ttfamily 1803.05088}}].

\bibitem{Schoorlemmer:2015afa}
H.~Schoorlemmer et~al., \emph{{Energy and Flux Measurements of Ultra-High
  Energy Cosmic Rays Observed During the First ANITA Flight}},
  \href{https://doi.org/10.1016/j.astropartphys.2016.01.001}{\emph{Astropart.
  Phys.} {\bfseries 77} (2016) 32}
  [\href{https://arxiv.org/abs/1506.05396}{{\ttfamily 1506.05396}}].

\bibitem{PhysRevLett.116.141103}
{\scshape T-510 Collaboration} collaboration, \emph{Accelerator measurements of
  magnetically induced radio emission from particle cascades with applications
  to cosmic-ray air showers},
  \href{https://doi.org/10.1103/PhysRevLett.116.141103}{\emph{Phys. Rev. Lett.}
  {\bfseries 116} (2016) 141103}.

\bibitem{paudel2022simulation}
E.~N. Paudel, A.~Coleman and F.~G. Schroeder, \emph{{Simulation study of the
  relative Askaryan fraction at the south pole}},
  \href{https://doi.org/10.1103/PhysRevD.105.103006}{\emph{Phys. Rev. D}
  {\bfseries 105} (2022) 103006}
  [\href{https://arxiv.org/abs/2201.03405}{{\ttfamily 2201.03405}}].

\bibitem{moliere1948theorie}
G.~Moliere, \emph{Theorie der streuung schneller geladener teilchen ii
  mehrfach-und vielfachstreuung},
  \href{https://doi.org/https://doi.org/10.1515/zna-1947-0302}{\emph{Zeitschrift
  f{\"u}r Naturforschung A} {\bfseries 3} (1948) 78}.

\end{thebibliography}\endgroup

\end{document}